\documentclass[11pt]{article}
\usepackage{graphicx}

\def \bx{{\bf x}}

\def \bp{{\bf p}}

\def \be{\begin{equation}}
\def \ee{\end{equation}}
\def \d{\partial}
\def \aa{\alpha}
\def \bb{\beta}
\def \dd{{\delta}}
\def \gg{\gamma}
\def \ss{\sigma}
\def \ll{\lambda}

\def \th{\theta}

\def \th{\theta}
\def \eps{\epsilon}
\def \kk{\kappa}
\def \DD{\Delta}

\def \GG{\Gamma}
\def \Om{\Omega}
\def \Th{\Theta}
\def \Ab{{\bar A}}

\def \Lc{{\cal L}}
\def \Nc{{\cal N}}
\def \Rc{{\cal R}}

\def \fr{\frac}

\def \veps{{\varepsilon}}

\def \ub{{\bar u}}

\def \pht{{\hat p}}
\def \xht{{\hat x}}
\def \dht{{\hat \d}}
\def \Aht{{\hat A}}
\def \Fht{{\hat F}}
\def \Jht{{\hat J}}
\def \Pht{{\hat P}}
\def \mb{{\bar m}}
\def \aab{{\bar \aa}}

\def \Ab{{\bar A}}

\def \Qb{{\bar Q}}

\def \psib{{\bar \psi}}

\def \ssb{{\bar \ss}}
\def \phib{{\bar \phi}}

\def \Tt{{\tilde{T}}} 
\def \Mt{{\tilde{M}}}
\def \b1{{\bf 1}}
\def \rt2{{\sqrt{2}}}

\begin{document}
\title{ Supersymmetry with Lorentz Symmetry Violation}
\author{I.T. Drummond\thanks{email: itd@damtp.cam.ac.uk} \\
            Department of Applied Mathematics and Theoretical Physics\\
            Centre for Mathematical Sciences\\
            Wilberforce Road\\ Cambridge\\ England, CB3 0WA
        }
\maketitle

\abstract{
	We study two (massless free field) models, a photon/photino model 
with a vector gauge field and a Majorana spinor field, and a Wess-Zumino 
model. They each exhibit Lorentz symmetry violation but retain, in an appropriate 
way, the supersymmetry correspondance between the particles of the two
fields. In relation to the photon field the Lorentz symmetry violation is of a
simple but non-trivial kind that implies birefringence. In relation to
the spinor field the Lorentz violation is produced by a modification of
the Majorana equation that is a simplified version of more general 
investigations of Lorentz symmetry violation of the Dirac equation.
In the case of the Wess-Zumino model we retain the same violation of Lorentz
symmetry for the Majorana field and adjust the propagation of the scalar
particles so that they exhibit a corresponding birefringence.
The advantages of the models are that they are straightforward to investigate
completely and both retain the basic aspect of supersymmetry namely
the one-to-one correspondance between bosons and fermions. As a result of this
bottom-up approach it is then possible to construct conserved supersymmetry charges
and investigate their algebraic properties. To some extent these are
similar to those encountered in the case of Lorentz invariance. However there are
differences and in particular non-local terms appear in the commutation relations
of the supersymmetry charges and fields of the models. We examine carefully
the rather intricate nature of the limit  back to Lorentz invariance.

\vfill
\pagebreak

\section{\bf Introduction}

Generally, and for good reasons, supersymmetry is viewed as fitting into the standard scheme of
Lorentz invariant field theory. However there are also good reasons for investigating the 
possibility that Lorentz symmetry might be violated and in particular at high energy \cite{VIEIRA,EICH1,EICH2,ITD5}.
One of the attractions of supersymmetry is that it can modify the high energy behaviour
of field theories rendering them strongly renormalisable or even finite. Hence
the possibility of retaining supersymmetry in a context in which Lorentz symmetry is
violated is to be taken seriously. There have been a number of proposals formulating models 
that retain supersymmetry while admitting Lorentz symmetry violation (LSV) \cite{KOST8,NIBPOS,COLL4,AYP1,AYP2}. 
A way of understanding some of the of issues involved is to adopt a distinction
between {\it intrinsic} and {\it extrinsic} Lorentz symmetry violation. An example is LSV in
deep inelastic scattering of electrons on hadrons \cite{KOST11,SHER1,KOST12,SHERRILL}. If we assume that the electron beam is 
as it is usually understood to be then LSV effects might be due either to a modification of the Dirac equation
describing the quarks making up the hadrons (intrinsic LSV), or to a distortion of the 
spacetime relationship between the quarks, perfectly standard  in their own frame that differs however from that
of the elctron beam (extrinsic LSV) (see the discussions in refs \cite{COLL4,SHERRILL}). Of course both types of LSV together 
with additional spin modification of the electron beam may  be present.

We regard the case of intrinsic LSV as of particular interest and  propose two models
to illustrate this. The first comprises a vector field, the "photon", and a massless Majorana spinor 
field, the "photino". The Lorentz symmetry breaking is unequivocal, both particles
exhibit birefringence in the form of a double lightcone. The parameters in the model can be adjusted
so that the dispersion relations for the photon and the photino conform appropriately with one another.
This makes it possible to make a pairing, for each lightcone, between a photon state of given 
4-momentum and a photino state with the same 4-momentum. We are then able to construct conserved 
supersymmetry charges that convert one type of particle into the other, thus justifying the photon/photino nomenclature.
However although the supersymmetry charges are obtained from locally conserved currents they are
more limited in scope than the standard supersymmetry charges of the fully Lorentz invariant model.
For example, they cannot connect states on distinct lightcones. 

The second model is a simplified Wess-Zumino model \cite{WESSZUM} in which we retain the massless Majorana particle and 
introduce scalar and pseudoscalar particles that travel on lightcones that match the
birefringent lightcones of the Majorana particle. In a manner similar to that of the photon/photino 
model we find conserved supersymmetry charges that are again of restricted scope. In both models
this feature complicates the approach to Lorentz symmetry where the full supersymmetry should be recovered. 
We examine this limiting procedure in some detail.

\section{Photon Lagrangian}

On setting up the photon/photino model we introduce, on the basis of discussions of Lorentz symmetry breaking 
in refs \cite{KOST2,SCHR1,ITD5,ITD3} for the vector field of the photon, $A_{\mu}(x)$, the Lagrangian 
\be
\Lc_P(x)=-\fr{1}{4}F^{\mu\nu}(x)F_{\mu\nu}(x)+\fr{1}{8}C^{\mu\nu\ll\tau}F_{\mu\nu}(x)F_{\ll\tau}(x).
\label{PLAG1}
\ee
Here $F_{\mu\nu}(x)=\d_{\mu}A_{\nu}(x)-\d_{\nu}A_{\mu}(x)$, and $C^{\mu\nu\ll\tau}$ is a tensor
with the algebraic symmetries of a Weyl tensor namely
\be
C^{\mu\nu\ll\tau}=C^{\ll\tau\mu\nu}=-C^{\nu\mu\ll\tau},
\ee
together with a zero-trace condition
\be
\eta_{\mu\ll}C^{\mu\nu\ll\tau}=0,
\ee
and a Bianchi identity
\be
C^{\mu\nu\ll\tau}+C^{\mu\ll\tau\nu}+C^{\mu\tau\mu\ll}=0.
\ee
Here $\eta_{\mu\ll}$ is the diagonal metric tensor with entries $(1,-1,-1,-1)$.
This framework covers all possible types of (intrinsic) Lorentz symmetry violation for the photon field.
From the Lagrangian $\Lc_P(x)$ we obtain the equation of motion
\be
\d_{\mu}G^{\mu\nu}(x)=0,
\label{PLAG1A}
\ee
where
\be
G^{\mu\nu}=-\fr{\d\Lc_P}{\d(\d_{\mu}A_{\nu}(x))}=F^{\mu\nu}(x)-C^{\mu\nu\ll\tau}\d_{\ll}A_{\tau}(x).
\label{PLAG1B}
\ee

Petrov \cite{PTRV} identified a classification of the possible versions of $C^{\mu\nu\ll\tau}$.
The Petrov classification is based on the principal null vectors of the Weyl tensor \cite{PENRIN}. 
In the general case there are four for a given Weyl tensor. However they may coincide in various ways 
and this is the basis of the Petrov scheme. In our model we select the simplest case, 
class N, in which they all coincide. If we denote this principal null vector by $l_{\mu}$ we can include 
it in a Penrose null tetrad \cite{PENRIN} comprising vectors $l_{\mu}$, $n_{\mu}$, $m_{\mu}$, $\mb_{\mu}$, that satisfy 
\be
l^2=n^2=m^2=\mb^2=l.m=l.\mb=n.m=n.\mb=0,
\ee
together with
\be
l.n=-m.\mb=1.
\ee
Explicitly, we choose $l^{\mu}=(1/\rt2,0,0,1/\rt2)$, $n^{\mu}=(1/\rt2,0,0,-1/\rt2)$,
$m^{\mu}=(0,1/\rt2,i/\rt2,0)$, $\mb=(0,1/\rt2,-i/\rt2,0)$. We can complete the construction of 
the photon Lagrangian by setting (see \cite{PENRIN})
\be
C^{\mu\nu\ll\tau}=\kk(A^{\mu\nu}A^{\ll\tau}+\Ab^{\mu\nu}\Ab^{\ll\tau}),
\label{LVINT1}
\ee
where
\be
A^{\mu\nu}=l^{\mu}m^{\nu}-l^{\nu}m^{\mu}
\ee
and
\be
\Ab^{\mu\nu}=l^{\mu}\mb^{\nu}-l^{\nu}\mb^{\mu}
\ee
An important property of the bi-vector $A^{\mu\nu}$ is self-duality, that is
\be
\fr{i}{2}\eps^{\mu\nu\ll\tau}A_{\ll\tau}=A^{\mu\nu},
\ee
while the bi-vector $\Ab^{\mu\nu}$ is anti-self-dual, that is
\be
\fr{i}{2}\eps^{\mu\nu\ll\tau}\Ab_{\ll\tau}=-\Ab^{\mu\nu}.
\ee
It is useful to note that
\be
l_{\mu}A^{\mu\nu}=l_{\mu}\Ab^{\mu\nu}=0,
\ee
and
\be
A^{\mu\nu}\Ab_{\mu\ll}=-l^{\nu}l_{\ll}.
\ee

\subsection{Photon Dynamics}

Because of the presence of the Lorentz symmetry violating term in
eq(\ref{PLAG1}) the model has unconventional features, we present the 
quantisation of the photon field in some detail. Following refs \cite{COLL1,COLL2,SCHR1}
we adopt the well known Gupta-Bleuler method adapted to our new circumstances. The first step is
to replace the Lagrangian in eq(\ref{PLAG1}) with
\be
\Lc_{GB}(x)=-\fr{1}{2}\d_{\mu}A_{\nu}\d^{\mu}A^{\nu}+\fr{1}{2}C^{\mu\nu\ll\tau}\d_{\mu}A_{\nu}\d_{\ll}A_{\tau}.
\label{PLAG2}
\ee
This is equivalent to the original version (up to total derivative terms) for fields
obeying the gauge constraint $\d.A(x)=0$. The equation of motion is 
\be
\d_{\mu}\left(\fr{\d\Lc_{GB}}{\d(\d_{\mu}A_{\nu})}\right)=0.
\label{EQMOT1}
\ee
This becomes
\be
-\d^2A^{\nu}+\d_{\mu}C^{\mu\nu\ll\tau}\d_{\ll}A_{\tau}=0.
\label{EQMOT2}
\ee
The Hamiltonian formulation of the photon dynamics requires the construction of the field $\Pi^{\mu}(x)$
canonically conjugate to $A_{\mu}(x)$. This is given by 
\be
\Pi^{\nu}(x)=\fr{\d \Lc_P}{\d(\d_{0} A_{\nu}(x))}=-K^{\nu\tau}{\dot A}_{\tau}(x)+\sum_{k=1}^{3}C^{0\nu k\tau}\d_k A_{\tau}(x),
\label{EQMOT3}
\ee
where
\be
K^{\nu\tau}=\eta^{\nu\tau}-C^{0\nu 0\tau}.
\label{EQMOT4}
\ee

\subsection{Plane Wave Solutions}

Plane wave solutions have the form
\be
A^{\nu}(x)=\veps^{\nu}e^{-ip.x}.
\ee
From eq(\ref{EQMOT2}) we have
\be
p^2\veps^{\nu}-p_{\mu}p_{\ll}C^{\mu\nu\ll\tau}\veps_{\tau}=0.
\label{EQMOT5}
\ee
The gauge condition is 
\be 
p.\veps=0.
\ee
The wave equation becomes
\be
p^2\veps^{\nu}-\kk(p_{\mu}A^{\mu\nu}p_{\ll}A^{\ll\tau}+p_{\mu}\Ab^{\mu\nu}p_{\ll}\Ab^{\ll\tau})\veps_{\tau}=0.
\label{EQMOT6}
\ee
There are four positive energy solutions. Two are "unphysical", comprising a gauge solution, $\veps_G^{\nu}=p^{\nu}$ and a complementary
solution $\veps_C^{\nu}=l^{\nu}$. Both these solutions require $p^2=0$ which implies that the mass-shell cone
and the lightcone on which they propagate are the standard cones associated with the metric $\eta^{\mu\nu}$.
There are also two "physical" solutions for which $p^2\ne 0$. It is immediately obvious from eq(\ref{EQMOT6})
any such solution has the form
\be
\veps^{\nu}=\aa p_{\mu}A^{\mu\nu}+\aab p_{\mu}\Ab^{\mu\nu},
\ee
for some values of $\aa$ and $\aab$.
On substituting this form into eq(\ref{EQMOT6}) we obtain
\begin{eqnarray}
	p^2\aa+\kk(l.p)^2\aab&=&0,\\
	p^2\aab+\kk(l.p)^2\aa&=&0.
\label{PMODES1}
\end{eqnarray}	
To obtain a nontrivial solution we require
\be
(p^2)^2-\kk^2(l.p)^4=0.
\label{PMODES2}
\ee
That is
\be
p^2\pm \kk(l.p)^2=0.
\label{PMODES3}
\ee

For simplicity of presentation we assume that $\kk>0$. It is then obvious that the two mass-shell cones are nested, 
the "-"cone lying in the interior of the "+"cone, except where 
they touch along a common generator parallel to $l^{\mu}$. Indeed they share this generator with
the standard cone $p^2=0$ (appropriate to the unphysical solutions) which is nested between the physical "$\pm$"cones.
However it should be noted that when $\kk=2$ the "-"cone acquires a generator parallel to the 0-axis
and the "+"cone acquires a generator along the (negative) 3-axis. We impose the constraint $\kk<2$. We comment on the significance of 
this constraint later. When $\kk<0$ the "+" and "-" cones interchange roles in the nesting structure. For this reason we impose 
also the corresponding constraint $\kk>-2$. Ultimately then we have (see also ref \cite{KLINK1})
\be
-2<\kk<2.
\label{PMODES4}
\ee
Subject to this restriction we can identify four positive energy solutions. The negative energy solutions 
are obtained by complex conjugation. From eq(\ref{PMODES3}) we find the allowed momenta 
$p^{\mu}_{\pm}=(E_{\pm},\bp)=(E_{\pm},p_1,p_2,p_3)$ where
\be
E_{\pm}=\fr{1}{1\pm\kk/2}\left(\pm(\kk/2)p_3+\sqrt{(1\pm\kk/2)(p_1^2+p_2^2)+p_3^2}\right).
\label{PMODES5}
\ee
Taking account of the relevant mass-shell conditions, the solutions to eqs(\ref{PMODES1}) are $\aa_{\pm}=\pm\aab_{\pm}$. 
The physical solutions, conveniently normalised, are then
\be
\veps^{\mu}_{+}(p_{+})=(l.p_{+})e^{\mu}_1-(e_1.p_{+})l^{\mu},
\label{PMODES6}
\ee
and
\be
\veps^{\mu}_{-}(p_{-})=(l.p_{-})e^{\mu}_2-(e_2.p_{-})l^{\mu},
\label{PMODES7}
\ee
where $e^{\mu}_1=(0,1,0,0)$ and $e^{\mu}_2=(0,0,1,0)$.

\subsection{\label{LCSTR} Lightcone Structure}

It is useful to examine the lightcone behaviour of the photons since this governs the causal
structure of the model. Consider the "+"mode. Define a transform of the momentum $p_{+}\rightarrow \pht_{+}$
where
\be
\pht_{+}^{\mu}=p_{+}^{\mu}+\fr{\kk}{2}l^{\mu}(l.p_{+}).
\label{LCONE1}
\ee
This mapping is a null shear in momentum space.
It is immediately obvious from eq(\ref{PMODES3}) that
\be
\pht_{+}^2=p_{+}^2+\kk(l.p_{+})^2=0.
\label{LCONE2}
\ee
The null shear maps the "+"mass-shell cone onto the standard null cone.
Similarly in spacetime introduce $x\rightarrow \xht_{+}$ where
\be
\xht_{+}^{\mu}=x^{\mu}-\fr{\kk}{2}l^{\mu}(l.x).
\label{LCONE3}
\ee
This is a null shear (of the opposite sign) in spacetime.  
We have immediately
\be
\xht_{+}.\pht_{+}=x.p_{+},
\label{LCONE4}
\ee
and therefore the phase factor $f=\exp\{-ix.p_{+}\}=\exp\{-i\xht_{+}.\pht_{+}\}$
satisfies the wave equation
\be
\dht_{+}^2 f=-\pht_{+}^2 f=0.
\label{LCONE5}
\ee
Here we have set
\be
\dht_{+\mu}=\fr{\d}{\d\xht_{+}^{\mu}}.
\label{LCONE6}
\ee
It follows that in terms of the coordinates $\xht_+^{\mu}$ the "+"lightcone is $\xht_+^2=0$,
which yields
\be
x^2-\kk(l.x)^2=\xht_+^2=0.
\label{LCONE7}
\ee
A similar discussion for the "-"mode leads to the transformation
\be
p_{-}^{\mu}\rightarrow \pht_{-}^{\mu}=p_{-}^{\mu}-\fr{\kk}{2}l^{\mu}(l.p_{-})
\label{LCONE8},
\ee
and
\be
x^{\mu}\rightarrow \xht_{-}^{\mu}=x^{\mu}+\fr{\kk}{2}l^{\mu}(l.x),
\label{LCONE9}
\ee
and the conclusion that the corresponding lightcone is 
\be
\xht_{-}^2=x^2+\kk (l.x)^2=0.
\label{LCONE10}
\ee
The lightcone structure therefore is similar, to that of the mass-shells in momentum space, except that now the interior
cone, the {\it slow} cone, is the "+"cone and the exterior cone, the {\it fast} cone, is the "-"cone.  

A geometrical understanding of the need for a restriction on the range of $\kk$ can be obtained by noting that 
when $\kk=2$ the "+"cone tilts so that it acquires a generator along the time axis. When $\kk>2$ the positive time
"+"cone lies entirely within the region $x_3>0$ at which point the co-ordinate system represented
by $x^{\mu}=(t,\bx)=(t,x_1,x_2,x_3)$ is no longer appropriate for describing the causal evolution of the model.
This picture can be developed further by noting that observers each associated with a reference frame that
has $\eta^{\mu\nu}$ as a metric, have coordinate systems related by Lorentz transformations. 
One such transformation is $L(\psi)$ a  boost along the (negative) 3-axis of velocity $v=\tanh\psi$.
It is easy to check that under such a transformation 
\be
l^{\mu}\rightarrow L^{\mu}_{~~\nu}(\psi)l^{\nu}=e^{\psi}l^{\mu}.
\label{LCONE11}
\ee
The description of the model in the new frame is unchanged provided we make the replacement
$\kk\rightarrow \kk'=e^{2\psi}\kk$. Even if $\kk$ lies within the acceptable range a
sufficiently powerful boost will shift $\kk'$ out of this range. A boost in the opposite direction
can be of any strength and will cause a shift eventually to observers who, because of limitations of measurement
accuracy, are unable to detect the Lorentz symmetry violation. The constraint on $\kk$ can be
reinterpreted as a constraint on the allowed reference frames. Observers may not be boosted to the
point at which they can overtake particles moving on the slow lightcone. We develop this point further in
section \ref{CONC}.

\subsection{\label{OVLPSEM} Overlaps of the Photon Wavefunctions}

In order to control the normalisation of wavefunctions it is necessary to define a scalar product
or overlap between them. We follow ref\cite{COLL1} and define the overlap of two solutions, $A^{\mu}(x)$ 
and $B^{\mu}(x)$, of eq(\ref{EQMOT2}) to be (A,B) where
\begin{eqnarray}
	(A,B)&=&-i\int d^3\bx \left(A^{*}_{\nu}(x)\d^0B^{\nu}(x)-B_{\nu}(x)\d^0A^{\nu*}(x)\right.\nonumber\\
	&&~~~~~~~~~~~~~~~\left.-C^{0\nu\ll\tau}(A^{*}_{\nu}(x)\d_{\ll}B_{\tau}(x)-B_{\nu}(x)\d_{\ll}A^{*}_{\tau}(x))\right).
\label{SCALP}									     
\end{eqnarray}									     
It is easy to verify that it is independent of time. We denote the wavefunctions as
\begin{eqnarray}
\Psi_{G\mu}(\bp,x)&=&p_{\mu}e^{-ip.x}\nonumber\\
\Psi_{C\mu}(\bp,x)&=&l_{\mu}e^{-ip.x}\nonumber\\
\Psi_{+\mu}(\bp,x)&=&((l.p_{+})e_{1\mu}-(e_1.p_{+})l_{\mu})e^{-ip_{+}.x}=\veps_{+\mu}(p_{+})e^{-ip_{+}.x}\nonumber\\
\Psi_{-\mu}(\bp,x)&=&((l.p_{-})e_{2\mu}-(e_2.p_{-})l_{\mu})e^{-ip_{-}.x}=\veps_{-\mu}(p_{-})e^{-ip_{-}.x}
\end{eqnarray}
The overlaps for the unphysical wave functions are
\be
(\Psi_{G}(\bp),\Psi_{G}(\bp')=(\Psi_{C}(\bp),\Psi_{C}(\bp'))=0.
\ee
\be
(\Psi_{G}(\bp),\Psi_{C}(\bp'))=-2E(l.p)(2\pi)^3\dd^{3}(\bp-\bp')
\ee
The overlaps between unphysical and physical wavefunctions all vanish.
The non-vanishing overlaps for the physical wavefunctions are
\begin{eqnarray}
	(\Psi_{+}(\bp),\Psi_{+}(\bp'))=(2\pi)^3\dd^{(3)}(\bp-\bp')(E_{+}-p_3)^2(E_{+}+\fr{\kk}{2}(E_{+}-p_3))\nonumber\\
	(\Psi_{-}(\bp),\Psi_{-}(\bp'))=(2\pi)^3\dd^{(3)}(\bp-\bp')(E_{-}-p_3)^2(E_{-}-\fr{\kk}{2}(E_{-}-p_3))
\end{eqnarray}
The structure of the overlaps of the unphysical wavefunctions is equivalent to
that of the standard Gupta-Bleuler formalism.  That is, they comprise a
subspace, in this case spanned by two sets of zero norm wavefunctions, that
contains both positive and negative norm wavefunctions. Ultimately this is why the
corresponding excitations do not contribute to the values of physical, that is
gauge invariant, quantities. This is particularly significant here because the unphysical
lightcone is distinct from the birefringent structure of the physical light
cones and any unphysical signal will be transported in manner distinct from the physical signals.

\subsection{\label{QUANT} Quantisation of the Photon Field}

The quantisation of the photon field is achieved by imposing the (equal time)
commutation relations 
\be
[\Pi^{\nu}(x),A_{\ll}(x')]=-i\dd^{\nu}_{\ll}\dd^{(3)}(\bx-\bx').
\ee
The field $\Pi^{\nu}(x)$ is defined in eq(\ref{EQMOT3}), see also \cite{COLL1,COLL2}. 

A convenient way of exploiting the canonical commutation relations is through
the identity
\be
[(f,A),A_{\mu}(x)]=f_{\mu}^{*}(x),
\label{IDENT1}
\ee
where $f_{\mu}(x)$ is any arbitrary photon wavefunction and $A_{\mu}(x)$ is the quantum photon field.
We can obtain this result by noting that from the definition in eq(\ref{SCALP}) we obtain
\be
(f,A)=i\int d^3\bx' f_{\nu}^{*}(x')\Pi^{\nu}(x')+\Rc,
\ee
where the remainder term $\Rc$ commutes (at equal times) with $A_{\mu}(x)$. The identity follows.

The quantum field $A_{\mu}(x)$ can be separated into a number of terms. They are
\be
A_{\mu}(x)=A_{+\mu}(x)+A_{-\mu}(x)+A_{U\mu}(x),
\ee
where
\be
A_{\pm \mu}(x)=\int d^3\bp\fr{1}{\Nc_{\pm}(\bp)}\left[a_{\pm}(\bp)\Psi_{\pm\mu}(\bp,x)+a^{\dag}_{\pm}(\bp)\Psi^{*}_{\pm\mu}(\bp,x)\right],
\label{QMODES1}
\ee
and $A_{U\mu}(x)$ contains the unphysical mode contributions. We include a normalising factor $1/\Nc_{\pm}(\bp)$
in eq(\ref{QMODES1}) in order to permit the imposition of the non-vanishing commutation relations in the form
\be
[a_{\pm}(\bp),a^{\dag}_{\pm}(\bp')]=(2\pi)^3\dd^{(3)}(\bp-\bp').
\ee
It follows, for example, that
\be
[a_{+}(\bp),A_{\mu}(x)]=\fr{1}{\Nc_{+}(\bp)}\Psi_{+\mu}^{*}(\bp,x).
\label{IDENT2}
\ee
We also have from the overlap calculation
\be
(\Psi_{+}(\bp),A)=\fr{1}{\Nc_{+}(\bp)}(E_{+}-p_3)^2(E_{+}+\fr{\kk}{2}(E_{+}-p_3))a_{+}(\bp).
\label{IDENT3}
\ee
If we choose $f_{\mu}(x)=\Psi_{+\mu}(\bp,x)$ in the identity eq(\ref{IDENT1})
then using eq(\ref{IDENT2})and eq(\ref{IDENT3}) we find
\be
[(\Psi_{+}(\bp),A),A_{+\mu}(x)]=\fr{1}{(\Nc_{+}(\bp))^2}(E_{+}-p_3)^2(E_{+}+\fr{\kk}{2}(E_{+}-p_3))\Psi_{+\mu}^{*}(\bp,x).
\ee
It follows that 
\be
\Nc_{+}(\bp)=\sqrt{(E_{+}-p_3)^2(E_{+}+\fr{\kk}{2}(E_{+}-p_3))}.
\label{NRM1}
\ee
By a parallel discussion we can show that
\be 
\Nc_{-}(\bp)=\sqrt{(E_{-}-p_3)^2(E_{-}-\fr{\kk}{2}(E_{-}-p_3))}.
\label{NRM2}
\ee

\subsection{\label{PHOTONEMT} Photon Energy-momentum Tensor}

The energy-momentum tensor for photons in the Gupta-Bleuler formalism can be
computed along conventional lines in the form
\be
\Th_{GB}^{\mu\nu}(x)=\fr{\d\Lc_{GB}}{\d(\d_{\mu}A_{\ll})}\d^{\nu}A_{\ll}-\eta^{\mu\nu}\Lc_{GB}.
\ee
If we define
\be
G^{\mu\nu}=-\fr{\d\Lc_{GB}}{\d(\d_{\mu}A_{\nu})},
\ee
then the the Lagrangian is
\be
\Lc_{GB}=\fr{1}{4}G^{\mu\nu}F_{\mu\nu},
\ee
and the equation of motion is
\be
\d_{\mu}G^{\mu\nu}=0.
\ee
The energy momentum tensor becomes
\be
\Th_{GB}^{\mu\nu}(x)=-G^{\mu\ll}(x)\d^{\nu}A_{\ll}(x)+\fr{1}{4}\eta^{\mu\nu}G^{\ll\tau}(x)F_{\ll\tau}(x).
\ee
It follows readily that
\be
\d_{\mu}\Th_{GB}^{\mu\nu}(x)=0.
\ee
This version of the energy-momentum tensor, just as in the Lorentz symmetric case, is
unsatisfactory as a physical quantity because it is not gauge invariant. The remedy
is the same also see \cite{ITZZ}. We introduce a correction
\be
\Th_{C}^{\mu\nu}(x)=G^{\mu\ll}\d_{\ll} A^{\nu}(x).
\ee
The physical energy-momentum tensor is then
\be
\Th^{\mu\nu}(x)=\Th_{GB}^{\mu\nu}(x)+\Th_{C}^{\mu\nu}(x).
\label{PHYSEMT1}
\ee
It is conserved
\be
\d_{\mu}\Th^{\mu\nu}(x)=0.
\ee
The 4-momentum operator is
\be
P^{\nu}=\int d^3\bx\Th^{0\nu}(x).
\ee
In terms of mode operators we have
\be
P^{\nu}=P_{+}^\nu+P_{-}^{\nu},
\ee
where
\be
P_{\pm}^{\nu}=\fr{1}{2}\int\fr{d^3\bp}{(2\pi)^3}(a_{\pm}^{\dag}(\bp)a_{\pm}(\bp)+a_{\pm}(\bp)a_{\pm}^{\dag}(\bp))p_{\pm}^{\nu}.
\label{EMOPS}
\ee
In the Lorentz symmetric case the procedure we have adopted in constructing $\Th^{\mu\nu}(x)$ also renders it
symmetrical. This is not true in the presence of Lorentz symmetry violation, the reason being that the generators of 
Lorentz transformations $L^{\mu\nu}$ are computed as
\be
L^{\mu\nu}=\int d^3\bx (x^{\mu}\Th^{0\nu}(x)-x^{\nu}\Th^{0\mu}(x)).
\ee
Now
\be
\d_{\ll}(x^{\mu}\Th^{\ll\nu}(x)-x^{\nu}\Th^{\ll}(x))=\Th^{\mu\nu}(x)-\Th^{\nu\mu}(x).
\ee
The absence of symmetry for $\Th^{\mu\nu}(x)$ then implies that the generators $L^{\mu\nu}$
are not time time-independent which is the case in our model. 

\section{\label{MAJOR} Majorana Spinor Field}

The Lorentz invariant Lagrangian, $\Lc_M(x)$for the Majorana field $\psi(x)$, is
\be
\Lc_M(x)=\fr{i}{2}\psib(x)\gg^{\mu}\d_{\mu}\psi(x).
\label{LINVM}
\ee
Here $\gg^{\mu}$ are the standard Dirac matrices appropriate to the metric $\eta^{\mu\nu}$.
We will follow \cite{KOST9} and adopt a chiral representation. The Majorana field
satisfies the massless Dirac equation,
\be
\gg.\d\psi(x)=0.
\label{DEQN}
\ee
The charge conjugation transformation is  $\psi\rightarrow \psi_C$ where
\be
\psi_C=C(\psib)^T.
\label{MCOND1}
\ee
The $\gg^{\mu}$ obey the conditions
\be
C\gg^{\mu}C^{-1}=-(\gg^{\mu})^{T},
\label{MCOND2}
\ee
and 
\begin{eqnarray}
	C^T=C^{\dag}=-C\nonumber\\
	C^2=-1
\end{eqnarray}
These properties are satisfied by the representation
\be
C=\left(\begin{array}{cc}
	i\ss^2&0\\
	0&-i\ss^2
\end{array}\right)
\label{CREP}
\ee
These properties imply that $\psi_C(x)$ also satisfies eq(\ref{DEQN}). We are then
free to impose the Majorana condition $\psi(x)=\psi_C(x)$ thus reducing the Majorana
field to the two independent components appropriate for a photino field.

\subsection{\label{LVMAJ} Lorentz Symmetry Violation and the Majorana Field}

In order to introduce Lorentz symmetry violation into the evolution of the 
Majorana field we follow the work of a number of authors \cite{KOST10,KOST9,SCHRR} who have considered
the implications of introducing Lorentz symmetry violation by means of generalisations
of the Dirac equation. In constructing our model we require only a simplified version of 
that approach applied to the Majorana equation. We adopt then the modified Majorana Lagrangian $\Lc_{M}$ where
\be
\Lc_{M}(x)=\fr{i}{2}\psib(x)\GG^{\mu}\d_{\mu}\psi(x),
\label{MODMAJ1}
\ee
where
\be
\GG^{\mu}=\gg^{\mu}+\fr{1}{2}T^{\mu}_{~~\aa\bb}\ss^{\aa\bb},
\label{MODMAJ2}
\ee
and $\ss^{\aa\bb}=(i/2)[\gg^{\aa},\gg^{\bb}]$. The modified Majorana equation is
\be
\GG.\d\psi(x)=0.
\label{MODMAJ3}
\ee
It is easy to check that $\psi_C(x)$ is also a solution of eq(\ref{MODMAJ3})
so we can indeed impose the condition $\psi_C(x)=\psi(x)$ on the solutions.
The field $\pi(x)$ canonically conjugate to $\psi(x)$ is 
\be
\pi(x)=i\psib(x)\GG^{0},
\ee
and the canonical equal time anticommutation relation is
\be
\{\psi_{\aa}(x),\pi_{\bb}(x')\}=i\dd_{\aa\bb}\dd^{(3)}(\bx-\bx'),
\label{ETAC1}
\ee
more succinctly
\be
\{\psi(x),\psib(x')\}\GG^{0}=\dd^{(3)}(\bx-\bx').
\label{ETAC2}
\ee
\subsection{\label{DISPR}Plane Waves and Dispersion Relation}

Crucial to constructing a supersymmetric model is arranging for a
concordance between the photon and photino dispersion relations \cite{KOST8}. To
investigate this we need the plane wave solutions of eq(\ref{MODMAJ3}).
These have the form
\be
\psi(\bp,x)=u e^{-ip.x},
\ee
where $p=(E,\bp)$ with $E>0$ and $u(\bp)$ is spinor with components $(u_1,u_2,u_3,u_4)$.
Eq(\ref{MODMAJ3}) implies
\be
Mu=0,
\label{SPINEQ1}
\ee
where
\be
M=\GG.p=\gg.p+\fr{1}{2}T_{\aa\bb}\ss^{\aa\bb},
\label{MMOP1}
\ee
and
\be
T_{\aa\bb}=p_{\mu}T^{\mu}_{~~\aa\bb}.
\label{MMOP2}
\ee
Following the reasoning in \cite{KOST10,KOST9,SCHRR} we introduce the dual tensor
\be
\Tt^{\aa\bb}=\fr{1}{2}\eps^{\aa\bb\ll\tau}T_{\ll\tau},
\label{MMOP3}
\ee
together with 
\be
T^{(\pm)}_{\aa\bb}=\fr{1}{2}(T_{\aa\bb}\pm i\Tt_{\aa\bb}),
\label{MMOP4}
\ee
the self-dual and anti-self-dual parts of $T_{\aa\bb}$.
It then follows (see \cite{KOST10,KOST9,SCHRR}) that $\DD(\bp)=\det M$ is given by
\be 
\DD(\bp)=(p^2)^2+(T^{(+)})^2(T^{(-)})^2+8V^{(+)}.V^{(-)},
\ee
where 
\be V^{(\pm)}_{\nu}=p^{\mu}T^{(\pm)}_{\mu\nu}.
\label{MMOP5}
\ee
The dispersion relation we require is 
\be
\DD(p)=0.
\label{DISPR1}
\ee
In order to fix the model precisely we must choose a specific form for $T^{(\pm)\mu}_{\aa\bb}$. 
In doing so it helps to recall that the photon model involved the null vector $l_{\mu}$ and the
self- and anti-self-dual tensors $A_{\aa\bb}$ and $\Ab_{\aa\bb}$. These suggest themselves
as candidates for the chiral structure we seek. Our initial proposal is then 
\be
T^{(+)\mu}_{\aa\bb}(\th)=\xi e^{i\th}l^{\mu}A_{\aa\bb},
\label{MMOP6}
\ee
together with the complex conjugate relation, $\th$ being real,
\be
T^{(-)\mu}_{\aa\bb}(\th)=\xi e^{-i\th}l^{\mu}\Ab_{\aa\bb}.
\label{MMOP7}
\ee
In turn this yields
\be
T^{\mu}_{\aa\bb}(\th)=\xi l^{\mu}(e^{i\th}A_{\aa\bb}+e^{-i\th}\Ab_{\aa\bb}).
\label{MMOP8}
\ee
If we make the replacements $m\rightarrow m'=e^{i\th}m$ and $\mb\rightarrow \mb'=e^{-i\th}\mb$
in the choice of Penrose tetrad the Majorana field we see that relative to the photon tetrad 
this represents a clockwise rotation in the (1,2)-plane about the 3-axis. However the extra generality
represented by the angle $\th$ is spurious. If we define 
\be
\GG^{\mu}(\th)=\gg^{\mu}+\fr{1}{2}T^{\mu}_{\aa\bb}(\th)\ss^{\aa\bb},
\ee
then we can easily show that
\be
\GG^{\mu}(\th)=X(\th)\GG^{\mu}X(\th),
\ee
where $\GG^{\mu}=\GG^{\mu}(\th=0)$ and
\be
X(\th)=e^{i\th/2}\fr{1+\gg_5}{2}+e^{-i\th/2}\fr{1-\gg_5}{2}
\ee
The Majorana Lagrangian becomes
\be
\Lc_{M}(x)=\fr{i}{2}\psib(x)X(\th)\GG.\d X(\th)\psi(x).
\ee
By means of the field transformations $X(\th)\psi(x)\rightarrow \psi(x)$ and (consistently) $\psib(x)X(\th)\rightarrow \psib(x)$
we have
\be
\Lc_{M}(x)=\fr{i}{2}\psib(x)\GG.\d\psi(x).
\label{MAJMAJL}
\ee
The implication is that we can choose any value of $\th$ without changing the model. 
For convenience we then choose $\th=0$ and replace eq(\ref{MMOP8}) with
\be
T^{\mu}_{\aa\bb}=\xi l^{\mu}(A_{\aa\bb}+\Ab_{\aa\bb}).
\label{MMOP9}
\ee
For our model then we easily see that $T^{(\pm)2}=0$ and find for the dispersion relation
\be
(p^2)^2-8\xi^2(l.p)^4=0.
\label{DISPR2}
\ee
This coincides with the the result in eq(\ref{PMODES2}) when $\kk^2=8\xi^2$. There are then two 
possibilities 
\be
\kk=\pm 2\rt2\xi.
\ee
In either case the birefringent mass-shell cone structure of the Majorana field corresponds exactly
with that of the birefringent photons though with a differing matching of states in the two cases. We make the choice 
$\xi=\kk/(2\rt2)$ (implying $\xi>0$) for simplicity of exposition. We will however deal with the case 
$\xi=-\kk/(2\rt2)$ later when considering the limiting case of Lorentz invariance. 
There are a number of approaches to
deriving the expression for $\DD(p)$ but it will re-emerge straightforwardly when we examine the 
explicit form of the spinor wavefunctions.

To obtain explicit plane wave solutions we follow refs\cite{KOST10,KOST9,SCHRR} and adopt the chiral representation
for the $\gg^{\mu}$ namely
\be
\gg^{\mu}=\left(\begin{array}{cc}
		  0&\ssb^{\mu}\\
		  \ss^{\mu}&0
\end{array} \right),
\ee
where $\ss^{\mu}=(\b1,\ss^1,\ss^2,\ss^3)$ and $\ssb^{\mu}=(\b1,-\ss^1,-\ss^2,-\ss^3)$,
$\ss^k$ $k=1,2,3$ being the standard Pauli matrices.
and
\be
\gg_5=\left(\begin{array}{cc}
	\b1&0\\
	0&-\b1
\end{array} \right)
\ee
It is a little simpler to deal with the modified version of eq(\ref{SPINEQ1}) 
\be
\Mt u=0,
\label{SPINEQ2}
\ee
where
\be
\Mt=\gg.p M=p^2+\fr{1}{2}T_{\aa\bb}\gg.p\ss^{\aa\bb}.
\label{SPINEQ3}
\ee
Using the well-known identity
\be
\gg^{\ll}\ss^{\aa\bb}=i(\eta^{\ll\aa}\gg^{\bb}-\eta^{\ll\bb}\gg^{\aa})-\eps^{\ll\aa\bb\tau}\gg_{\tau}\gg_5.
\label{IDY}
\ee
$\Mt$ can be put in the form
\be
\Mt=\left(\begin{array}{cc}
	p^2&2iV^{(-)}.\ssb\\
	2iV^{(+)}.\ss&p^2
\end{array}\right),
\label{SPINEQ4}
\ee
where $V^{(\pm)}_{\mu}$ are defined in eq(\ref{MMOP5}). We then obtain the  
equations for the spinor $u$
\begin{eqnarray}
	p^2u_1-i2\rt2\xi(l.p)(\mb.p)u_3&=&0,\nonumber\\
	p^2u_2+i2\rt2\xi(l.p)^2 u_3&=&0,\nonumber\\
	-i2\rt2\xi(l.p)^2 u_2+p^2u_3&=&0,\nonumber\\
	-i2\rt2\xi(l.p)(m.p)u_2+p^2u_4&=&0.
\end{eqnarray}
In order to yield a nontrivial solution the first and last of these equations show 
that $u_2$ and $u_3$ cannot both vanish. The other two equations therefore
require that
\be
\det\left(\begin{array}{cc}
	p^2&i2\rt2\xi(l.p)^2\\
	-i2\rt2\xi(l.p)^2&p^2
\end{array}\right)=0.
\ee
That is of course identical to the dispersion relation from eq(\ref{DISPR2}) and,
on imposing the the relation $\kk=2\rt2\xi$ (which we will asume from here on), the same as that from eq(\ref{PMODES2}).
The mass-shell "$\pm$"cones are identical between photon and photino. The plane wave solutions are
\be
\psi_{\pm}(\bp,x)=u_{\pm}(p_{\pm})e^{-ip_{\pm}.x},
\ee
where
\be
u_{\pm}(p_{\pm})=\left(\begin{array}{c}
	    -(\mb.p_{\pm})\\
	    (l.p_{\pm})\\
	    \mp i(l.p_{\pm})\\
	    \mp i(m.p_{\pm})
\end{array}\right).
\label{PLWAVE1}
\ee
The positive energy plane waves are
therefore
\be
\psi_{\pm}(\bp,x)=u_{\pm}(p_{\pm})e^{-ip_{\pm}.x}.
\ee
The negative energy plane waves are the charge conjugate wavefunctions
\be
\psi_{\pm C}(\bp,x)=C(\psib_{\pm}(\bp,x))^T=u_{\pm C}(p_{\pm})e^{ip_{\pm}.x},
\ee
where
\be
u_{\pm C}(p_{\pm})=C(\ub_{\pm}(p_{\pm})^T=\mp iu_{\pm}(p_{\pm}).
\ee 
For general 4-vector $q$ we write
\be
u_{\pm}(q)=\left(\begin{array}{c}
	-(\mb.q)\\
	(l.q)\\
	\mp i(l.q)\\
	\mp i(m.q)
\end{array}\right).
\ee
It is easily checked, in the chiral representation for $\gg$-matrices, that
\be
u_{\pm}(q)=\gg.q~u_{\pm}^{(0)},
\ee
where
\be
u_{\pm}^{(0)}=\left(\begin{array}{c}
	\mp i/\rt2\\
	0\\
	0\\
	1/\rt2
\end{array}\right).
\ee
It also obvious that 
\be
u_{\pm}(q+sl)=u_{\pm}(q),
\ee
for any value of the scalar $s$. We have then
\be
u_{\pm}(p_{\pm})=\gg.p_{\pm}u_{\pm}^{(0)}=\gg.\pht_{\pm}u_{\pm}^{(0)},
\ee
with the result
\be
\gg.\pht_{\pm}u_{\pm}(p_{\pm})=(\gg.\pht_{\pm})^2u_{\pm}^{(0)}=0.
\ee
The Majorana wavefunctions can be put in the form
\be
\psi_{\pm}(\bp,x)=i\gg.\dht_{\pm} u^{(0)}_{\pm}e^{-ip_{\pm}.x}
\label{REFSWV4}
\ee
and
\be
\psi_{\pm C}(\bp,x)=C(\psib_{\pm}(\bp,x))^T=-i\gg.\dht_{\pm} u^{(0)}_{\pm C}e^{ip_{\pm}.x},
\label{REFSWV5}
\ee
where
\be
u^{(0)}_{\pm C}=C(\ub^{(0)}_{\pm})^{T}.
\label{REFSWV6}
\ee
We then have the result
\be
\gg.\dht_{\pm}\psi_{\pm}(\bp,x)=\dht_{\pm}^2\psi_{\pm}(\bp,x)=0.
\ee
The complete Majorana field comprises a superposition of these plane wave solutions.
It can be split into two parts $\psi_{\pm}(x)$ each associated in the obvious way 
with the "$\pm$" lightcones. We can  write
\be
\psi(x)=\psi_{+}(x)+\psi_{-}(x),
\ee
where
\be
\gg.\dht_{\pm}\psi_{\pm}(x)=0.
\label{REFSWV7}
\ee

\subsection{\label{OVLAPSMAJ} Overlaps of the Majorana Wavefunctions}

If $\psi(x)$ and $\phi(x)$ are Majorana wavefunctions then the current $J^{\mu}(x)=\phib(x)\GG^{\mu}\psi(x)$
is conserved 
\be
\d_{\mu}J^{\mu}(x)=0.
\ee
It is then possible to define a time independent overlap $(\phi,\psi)$
\be
(\phi,\psi)=\int d^3\bx \phib(x)\GG^0\psi(x).
\ee
The non-vanishing overlaps between the plane wave solutions are easily computed as 
\be
(\psi_{\pm}(\bp),\psi_{\pm}(\bp'))=(2\pi)^3\dd^{(3)}(\bp-\bp')2(E_{\pm}-p_3)(E_{\pm}\pm\fr{\kk}{2}(E_{\pm}-p_3)).
\ee
The same holds true ($\psi\rightarrow \psi_C$)  for the charge conjugate wavefunctions.

\subsection{\label{QMAJ} Quantisation of the Majorana Field}

We introduce mode operators $b_{\pm}(\bp),b^{\dag}_{\pm}(\bp)$ for the Majorana field by expanding $\psi_{\pm}(x)$
in the form
\be
\psi_{\pm}(x)=\int\fr{d^3\bp}{(2\pi)^3}\fr{1}{N_{\pm}(\bp)}(b_{\pm}(\bp)e^{-ip_{\pm}.x}
                                  \mp ib^{\dag}_{\pm}(\bp)e^{ip_{\pm}.x})u_{\pm}(p_{\pm}).
\label{MAJFLD}				  
\ee
The factor $\mp i$ in the second term of the integrand renders the field even under charge conjugation.
The normalising factor $N_{\pm}(\bp)$ is chosen so that the non-vanishing anticommutation relations 
\be
\{b_{\pm}(\bp),b_{\pm}^{\dag}(\bp')\}=(2\pi)^3\dd^{(3)}(\bp-\bp'),
\ee
are consistent with the canonical anticommutation relations eq(\ref{ETAC2}).
It is easily checked that
\be
N_{\pm}(\bp)=\sqrt{2(E_{\pm}-p_3)(E_{\pm}\pm(\kk/2)(E_{\pm}-p_3))}.
\label{MAJNORM}
\ee

\subsection{\label{PHTNOEMT} Energy-momentum Tensor for the Majorana Field}}

The energy momentum tensor for the Majorana field in our model is $\Th^{\mu\nu}_{M}(x)$ where
\be
\Th^{\mu\nu}_{M}(x)=\fr{i}{2}\psib(x)\GG^{\mu}\d^{\nu}\psi(x)-\eta^{\mu\nu}\Lc_{M}(x).
\ee
When $\psi(x)$ satisfies the equations of motion the Lagrangian contribution vanishes
and 
\be
\d_{\mu}\Th^{\mu\nu}_{M}(x)=0.
\ee
The 4-momentum $P^{\nu}$ is then given by
\be
P^{\nu}=\int d^3\bx \Th^{0\nu}_{M}(x),
\ee
and is independent of time. Expressed in terms of mode operators we have 
\be
P^{\nu}=P_{+}^{\nu}+P_{-}^{\nu},
\ee
where
\be
P_{\pm}^{\nu}=\fr{1}{2}\int\fr{d^3\bp}{(2\pi)^3}p_{\pm}^{\nu}(b^{\dag}_{\pm}(\bp)b_{\pm}(\bp)-b_{\pm}(\bp)b^{\dag}_{\pm}(\bp)).
\ee
It is immediately obvious that the vacuum contributions of the Majorana spinors to the 4-momentum cancel the
corresponding contributions of the gauge fields. This must be necessarily the case if supersymmetry is
to be maintained in the Lorentz symmetry breaking model.

\section{\label{SSCHGS}Supersymmetry Charges}

The photon/photino model is potentially supersymmetric, even in the presence of Lorentz symmetry 
breaking, because it is possible to align the
birefringent mass-shell cones and lightcones of the photons and photinos. 
The crucial stage in completing the model is the construction of supersymmetry charges.
We show that this is indeed possible although with a somewhat unconventional approach forced
on us by the birefringence that expresses the Lorentz symmetry violation in the model.
The conventional procedure is to derive a conserved Noether current from a symmetry
of the Lagrangian. We reverse the procedure and postulate a current that we show to be conserved.
Subsequently we explore the algebra of conserved charges and their efficacy in connecting
photon and photino states. 

Guided by the conventional form of the current we postulate 
currents, one for each mass-shell cone, $\Jht_{\pm\pm}^{\mu}(x)$ where
\be
\Jht_{\pm\pm}^{\mu}(x)=\Fht_{\pm\ll\tau}(x)\ss^{\ll\tau}\gg^{\mu}\psi_{\pm}(x).
\label{SSCH1}
\ee
Here we are using
\be
\Fht_{\pm\mu\nu}(x)=\dht_{\pm\mu}\Aht_{\pm\nu}-\dht_{\pm\nu}\Aht_{\pm\mu}(x),
\ee
where (recall that $l.A_{\pm}(x)=0$)
\be
\Aht_{\pm\mu}(x)=(\dd^{\ll}_{\mu}\pm \fr{\kk}{2}l_{\mu}l^{\ll})A_{\pm\ll}(x)=A_{\pm\mu}(x).
\ee
It is then easy to see that 
\be
\dht_{\pm}.\Aht_{\pm}(x)=\d.A_{\pm}(x)=0.
\ee
and that
\be
\dht_{\pm}^{\mu}\Fht_{\pm\mu\nu}(x)=\dht_{\pm}^2A_{\pm\nu}(x)=0.
\ee
We have also the Bianchi identity
\be
\dht_{\pm\ll}\Fht_{\pm\mu\nu}(x)+\dht_{\pm\mu}\Fht_{\pm\nu\ll}(x)+\dht_{\pm\nu}\Fht_{\pm\ll\mu}(x)=0.
\ee
Using the identity in eq(\ref{IDY}) and the Bianchi identity we see by a standard argument that
\be
\dht_{\pm\mu}\Fht_{\pm\ll\tau}(x)\ss^{\ll\tau}\gg^{\mu}=0.
\ee
Also from eq(\ref{REFSWV4}) we have
\be
\gg.\dht_{\pm}\psi_{\pm}(x)=0.
\ee
It follows immediately that
\be
\dht_{\pm\mu}\Jht_{\pm\pm}^{\mu}(x)=0.
\label{CNSRV1}
\ee
Introducing 
\be
J_{\pm\pm}^{\mu}(x)=\Jht_{\pm\pm}^{\mu}(x)\pm \fr{\kk}{2}l^{\mu}l.\Jht_{\pm\pm}(x),
\label{SCURTS}
\ee
we find
\be
\d_{\mu}J_{\pm\pm}^{\mu}(x)=0.
\ee

It is worth noting that we can also introduce the currents $\Jht_{\pm\mp}^{\mu}(x)$ where
\be
\Jht_{\pm\mp}^{\mu}(x)=\Fht_{\pm\ll\tau}(x)\ss^{\ll\tau}\gg^{\mu}\psi_{\mp}(x).
\label{SSCH2}
\ee
However these currents, for the present choice of $\xi=+\kk/(2\rt2)$, are not conserved.
For example we can easily show that
\be
\dht_{+\mu}\Jht_{+-}^{\mu}(x)=\kk \Fht_{+\ll\tau}(x)\ss^{\ll\tau}l.\gg l.\d\psi_{-}(x)\ne 0.
\label{CNSRV2}
\ee
The corresponding supersymmetry charges are therefore not independent of time and
reflect the presence of Lorentz symmetry violation in the model  when $\xi=+\kk/(2\rt2)$.
Suitably interpreted however when $\xi=-\kk/(2\rt2)$, they do provide conserved charges
because of the interchange of photino mass-shells that occurs in this case.
We consider this possibility  later.

The conserved supersymmetry charges with which we are concerned are given by
\be
Q_{\pm\pm}=\int d^3\bx J_{\pm\pm}^{0}(x).
\ee
When expressed in terms of the mode operators we have
\be
Q_{++}=2\int\fr{d^3\bp}{(2\pi)^3}\fr{1}{\sqrt{(E_{+}-p_3)}}(a_{+}(\bp)b^{\dag}_{+}(\bp)-ia^{\dag}_{+}(\bp)b_{+}(\bp))u_{+}(p_{+}),
\label{SSCH3}
\ee
and
\be
Q_{--}=2\int\fr{d^3\bp}{(2\pi)^3}\fr{1}{\sqrt{(E_{-}-p_3)}}(ia_{-}(\bp)b^{\dag}_{-}(\bp)-a^{\dag}_{-}(\bp)b_{-}(\bp))u_{+}(p_{-}).
\label{SSCH4}
\ee
We draw attention to the apparently anomalous factor $u_{+}(p_{-})$ in the integrand in eq(\ref{SSCH4}). This factor can be 
re-expressed as $\gg_5 u_{-}(p_{-})$ which leaves it more seemingly natural but requires the explicit presence of $\gg_5$.
We leave eq(\ref{SSCH4}) as it stands. We define also the conjugate charges
\be
\Qb_{\pm\pm}=Q_{\pm\pm}^{\dag}\gg^{0}.
\ee
We have immediately
\be
\{Q_{\pm\pm},\Qb_{\mp\mp}\}=0.
\ee
The  nonvanishing anticommutators are
\be
\{Q_{\pm\pm},\Qb_{\pm\pm}\}=8\int \fr{d^3\bp}{(2\pi)^3}\fr{1}{(E_{\pm}-p_3))}
        (a_{\pm}^{\dag}(\bp)a_{\pm}(\bp)+b^{\dag}_{\pm}(\bp)b_{\pm}(\bp))u_{+}(p_{\pm})\ub_{+}(p_{\pm}).
\label{ANTICOM1}	
\ee
using the result (which can be checked by taking the trace with a complete basis of $\gg$-matrices)
\be
u_{+}(p_{\pm})\ub_{+}(p_{\pm})=
         \fr{1}{2}\left[(E_{\pm}-p_3) \pht_{\pm}.\gg+\rt2\veps^{\ll}_{+}(p_{\pm})\pht^{\tau}_{\pm}\ss_{\ll\tau}
\right].
\label{WVFNID}
\ee								      
We can then show that								      
\be
\{Q_{\pm\pm},\Qb_{\pm\pm}\}=4\int\fr{d^3\bp}{(2\pi)^3}(a^{\dag}_{\pm}(\bp)a_{\pm}(\bp)+b^{\dag}_{\pm}(\bp)b_{\pm}(\bp))
  \left[\pht_{\pm}.\gg+\rt2\fr{\veps^{\ll}_{+}(p_{\pm})\pht_{\pm}^{\tau}\ss_{\ll\tau}}{(E_{\pm}-p_3)}
			  \right].
\label{ANTICOM3}      
\ee
The first term in the integrand in eq(\ref{ANTICOM3}) yields a contribution to the anticommutator
of the supersymmetry charges of the form
\be
\{Q_{\pm\pm},\Qb_{\pm\pm}\}=4\Pht_{\pm}.\gg+\cdots ,
\ee
where $\Pht_{\pm}$ is the appropriately modified 4-momentum operator for the photon/photino system.
This term is a contribution to the anticommutator similar to the standard result
for the Lorentz symmetric case and to which it reduces when $\xi\rightarrow 0+$. The remaining term seems
to stand in the way of reproducing the standard result in the limit $\xi\rightarrow 0$. 
However an examination of the corresponding limit $\xi\rightarrow 0-$ provides the appropriate
cancelling contributions.

\subsection{\label{INTCGCNS} Interchange of Mass-shell Cones}

The result of setting $\xi=-\kk/(2\rt2)$ is not only directly interesting but is
crucial in understanding the limit $\kk\rightarrow 0$ when Lorentz invariance is restored.
We retain the $\pm$ identification of the mass-shells extablished by the photon field.
The reversal of sign for $\xi$ interchanges the mass-shells for the photino field
with the outcome that
\be
\psi_{\mp}(x)=\int\fr{d^3\bp}{(2\pi)^3}\fr{1}{N_{\pm}(\bp)}
                     (b_{\mp}(\bp)e^{-ip_{\pm}.x}\pm ib^{\dag}_{\mp}(\bp)e^{ip_{\pm}.x})u_{\mp}(p_{\pm}).
		     \label{INTRCHa}
\ee
The time independent supersymmetry charges $Q_{+-}$ and $Q_{-+}$ can be constructed from the 
now conserved currents $J^{\mu}_{+-}(x)$ and $J^{\mu}_{-+}(x)$ where
\be
J^{\mu}_{\pm\mp}(x)=\Jht^{\mu}_{\pm\mp}(x)\pm(\kk/2)l^{\mu}l.\Jht_{\pm\mp}(x),
\ee
and
\be
\Jht^{\mu}_{\pm\mp}(x)=\Fht_{\pm\ll\tau}(x)\ss^{\ll\tau}\gg^{\mu}\psi_{\mp}(x).
\ee
The supersymmetry charges take the form
\be
Q_{+-}=2\int\fr{d^3\bp}{(2\pi)^3}\fr{1}{\sqrt{(E_{+}-p_3)}}(a_{+}(\bp)b^{\dag}_{-}(\bp)+ia^{\dag}_{+}(\bp)b_{-}(\bp))u_{-}(p_{+}).
\label{INTRCH1}
\ee
Similarly we have
\be
Q_{-+}=2\int\fr{d^3\bp}{(2\pi)^3}\fr{1}{\sqrt{(E_{-}-p_3)}}(ia_{-}(\bp)b^{\dag}_{+}(\bp)+a^{\dag}_{-}(\bp)b_{+}(\bp))u_{-}(p_{-}).
\label{INTRCH2}
\ee

Following the scheme of previous calculations we find the nonvanishing anticommutators to be
\be
\{Q_{\pm\mp},\Qb_{\pm\mp}\}=4\int\fr{d^3\bp}{(2\pi)^3}(a^{\dag}_{\pm}(\bp)a_{\pm}(\bp)+b^{\dag}_{\mp}(\bp)b_{\mp}(\bp))
		  \left[\pht_{\pm}.\gg-\rt2\fr{\veps^{\ll}_{+}(p_{\pm})\pht^{\tau}_{\pm}\ss_{\ll\tau}}{(E_{\pm}-p_3)}\right].
\label{ANTICOM4}		  
\ee
Note the change of sign for the second term in square brackets relative to the corresponding term in eq(\ref{ANTICOM3}).

\subsection{\label{LSYMLIM}Lorentz Invariant Limit}

Lorentz invariance is achieved in the model by setting $\xi=0$. We have then that $p_{\pm}$ and $\pht_{\pm}$ all reduce
to a common value of $p$ where $p^2=0$. We have also in this limit $\Nc_{+}(\bp)=\Nc_{-}(\bp)=\Nc(\bp)=\sqrt{(E-p_3)^2 E}$ and
$N_{+}(\bp)=N_{-}(\bp)=N(\bp)=\sqrt{2(E-p_3)E}$.
In that case the supersymmetry charges, all of which are constant,
take the form
\begin{eqnarray}
       Q_{++}&=&2\int\fr{d^3\bp}{(2\pi)^3}\fr{1}{\sqrt{(E-p_3)}}(a_{+}(\bp)b^{\dag}_{+}(\bp)
	                        - ia^{\dag}_{+}(\bp)b_{+}(\bp))u_{+}(p).\nonumber\\
       Q_{--}&=&2\int\fr{d^3\bp}{(2\pi)^3}\fr{1}{\sqrt{(E-p_3)}}(ia_{-}(\bp)b^{\dag}_{-}(\bp)
	                          -a^{\dag}_{-}(\bp)b_{-}(\bp))u_{+}(p).\nonumber\\
	Q_{+-}&=&2\int\fr{d^3\bp}{(2\pi)^3}\fr{1}{\sqrt{(E-p_3)}}(a_{+}(\bp)b^{\dag}_{-}(\bp)
					+i a^{\dag}_{+}(\bp)b_{-}(\bp))u_{-}(p).\nonumber\\
       Q_{-+}&=&2\int\fr{d^3\bp}{(2\pi)^3}\fr{1}{\sqrt{(E-p_3)}}(ia_{-}(\bp)b^{\dag}_{+}(\bp)
                                         +a^{\dag}_{-}(\bp)b_{+}(\bp))u_{-}(p).
\label{LISSQ}
\end{eqnarray}
The full list of anticommutation relations is given in appendix \ref{ACOMS}. 
The diagonal relations can be read off from eqs(\ref{ANTICOM3}) and (\ref{ANTICOM4}) by setting
$\kk=0$. We have
\be
\{Q_{\pm\pm},\Qb_{\pm\pm}\}=4\int\fr{d^3\bp}{(2\pi)^3}\left[p.\gg+\rt2\fr{\veps^{\ll}_{+}(p)p^{\tau}\ss_{\ll\tau}}{E-p_3}\right]
		(a^{\dag}_{\pm}(\bp)a_{\pm}(\bp)+b^{\dag}_{\pm}(\bp)b_{\pm}(\bp)),
\label{ANTICOM5}
\ee
and
\be
\{Q_{\pm\mp},\Qb_{\pm\mp}\}=4\int\fr{d^3\bp}{(2\pi)^3}\left[p.\gg-\rt2\fr{\veps^{\ll}_{+}(p)p^{\tau}\ss_{\ll\tau}}{E-p_3}\right]
		(a^{\dag}_{\pm}(\bp)a_{\pm}(\bp)+b^{\dag}_{\mp}(\bp)b_{\mp}(\bp)).
\label{ANTICOM6}
\ee
Of the off-diagonal anticommutators some are are directly zero. The others as can be seen from appendix \ref{ACOMS},
yield contributions that cancel in pairs.
It follows that if we set
\be
Q=Q_{++}+Q_{--}+Q_{+-}+Q_{-+},
\label{TOTQ}
\ee
then we find
\be
\{Q,\Qb\}=8\int\fr{d^3\bp}{(2\pi)^3}p.\gg(a^{\dag}_{+}(\bp)a_{+}(\bp)+a^{\dag}_{-}(\bp)a_{-}(\bp)
		+b^{\dag}_{+}(\bp)b_{+}(\bp)+b^{\dag}_{-}(\bp)b_{-}(\bp)) = 8\gg.P,
\ee
where $P^{\mu}$ is the complete 4-momentum operator for the model in the Lorentz invariant limit.
This demonstrates that in the Lorentz invariant case, we can recover the 
complete constant supersymmetric charge with the correct anticommutation relation.
However it is evident that to achieve this outcome it is necessary to include contributions from {\it both} limits
$\xi\rightarrow 0\pm$.

\subsection{\label{QFCOMS} Action of Supersymmetry Charges on Modes and Fields}

We return to the broken symmetry case with $\xi=+\kk/(2\rt2)$.
The action of the supersymmetry charges on mode operators can be read off from
their definition in eq(\ref{SSCH3}) and eq(\ref{SSCH4}). We have
\begin{eqnarray}
	{[Q_{++},a_{+}(\bp)]}&=&\fr{ 2i}{\sqrt{(E_{+}-p_3)}}b_{+}(\bp)u_{+}(p_{+})\nonumber\\
	{[Q_{++},a^{\dag}_{+}(\bp)]}&=&\fr{2}{\sqrt{(E_{+}-p_3)}}b^{\dag}_{+}(\bp)u_{+}(p_{+})\nonumber\\
	{\{b_{+}(\bp),\Qb_{++}\}}&=&\fr{2i}{\sqrt{(E_{+}-p_3)}}a_{+}(\bp)\ub_{+}(p_{+})\nonumber\\
        {\{b^{\dag}_{+}(\bp),\Qb_{++}\}}&=&\fr{2}{\sqrt{(E_{+}-p_3)}}a^{\dag}_{+}(\bp)\ub_{+}(p_{+})
\label{MODCOMS1}			
\end{eqnarray}
and
\begin{eqnarray}
	{[Q_{--},a_{-}(\bp)]}&=&\fr{2}{\sqrt{(E_{-}-p_3)}}b_{-}(\bp)u_{+}(p_{-})\nonumber\\
        {[Q_{--},a^{\dag}_{-}(\bp)]}&=&\fr{2i}{\sqrt{(E_{-}-p_3)}}b^{\dag}_{-}(\bp)u_{+}(p_{-})\nonumber\\
        {\{b_{-}(\bp),\Qb_{--}\}}&=&\fr{-2}{\sqrt{(E_{-}-p_3)}}a_{-}(\bp)\ub_{+}(p_{-})\nonumber\\
        {\{b^{\dag}_{-}(\bp),\Qb_{--}\}}&=&\fr{-2i}{\sqrt{(E_{-}-p_3)}}a^{\dag}_{-}(\bp)\ub_{+}(p_{-})	
\label{MODCOMS2}                        
\end{eqnarray}
The (anti)commutators therefore convert the photon/photino mode operators correctly into
the corresponding photino/photon mode operators.
The action of the supersymmetry charges on fields can be deduced immediately.

\subsubsection{Anticommutators for Majorana fields}

From eq(\ref{MAJFLD}) and eq(\ref{MAJNORM}) we have
\be
\{\psi_{+}(x),\Qb_{++}\}=i\rt2\int\fr{d^3\bp}{(2\pi)^3}\fr{1}{\Nc_{+}(\bp)}(a_{+}(\bp)e^{-ip_{+}.x}-a^{\dag}_{+}(\bp)e^{ip_{+}.x})
			  u_{+}(p_{+})\ub_{+}(p_{+}).
\label{FLDCOMS1}			  
\ee
We have used the result
\be
N_{+}(\bp)(E_{+}-p_3)^{1/2}=\rt2\Nc_{+}(\bp).
\ee
From eq(\ref{WVFNID}) we conclude that right side of eq(\ref{FLDCOMS1}) becomes
$$
-\dht_{+\tau}\int\fr{d^3\bp}{(2\pi)^3}\fr{1}{\Nc_{+}(\bp)}(a_{+}(\bp)e^{-ip_{+}.x}+a^{\dag}_{+}(\bp)e^{ip_{+}.x})\left[\gg^{\tau}(l.p_{+})
                          +\veps_{+\ll}(p_{+})\ss^{\ll\tau}\right].
$$
Expressed in terms of fields eq(\ref{FLDCOMS1}) takes the form
\be
\{\psi_{+}(x),\Qb_{++}\}=\fr{1}{2}\Fht_{+\ll\tau}\ss^{\ll\tau}-\gg.\dht_{+}l.\d\phi_{+}(x),
\ee
where
\be
\phi_{+}(x)=i\int\fr{d^3\bp}{(2\pi)^3}\fr{1}{\Nc_{+}(\bp)}(a_{+}(\bp)e^{-ip_{+}.x}-a^{\dag}_{+}(\bp)e^{ip_{+}.x}).
\ee

Similarly we have
\be
\{\psi_{-}(x),\Qb_{--}\}=-\rt2\int\fr{d^3\bp}{(2\pi)^3}\fr{1}{\Nc_{-}(\bp)}(a_{-}(\bp)e^{-ip_{-}.x}-a^{\dag}_{-}e^{ip_{-}.x})
                                   u_{-}(p_{-})\ub_{+}(p_{-})
\ee
Making use of the identities
\be
\gg_5\ss^{\ll\tau}=\fr{i}{2}\eps^{\ll\tau\aa\bb}\ss_{\aa\bb},
\ee
and
\be
\veps_{+\ll}(p_{-})\pht_{-\tau}\gg_5\ss^{\ll\tau}=i\pht_{-\ll}\veps_{-\tau}(p_{-})\ss^{\ll\tau},
\ee
we can show that
\be
\{\psi_{-}(x),\Qb_{--}\}=\fr{1}{2}\Fht_{-\ll\tau}\ss^{\ll\tau}-i\gg_5\gg.\dht_{-} l.\d\phi_{-}(x),
\ee
where
\be
\phi_{-}(x)=i\int\fr{d^3\bp}{(2\pi)^3}\fr{1}{\Nc_{-}(\bp)}(a_{-}(\bp)e^{-ip_{-}.x}-a^{\dag}_{-}(\bp)e^{ip_{-}.x}).
\ee
These anticommutation relations for the the supersymmetry charges involve the 
newly introduced additional fields $\phi_{\pm}(x)$. Since they are built from the mode operators $a_{\pm}(\bp),a^{\dag}_{\pm}(\bp)$
they involve the same degrees of freedom as the original vector fields.

\subsubsection{Commutators for the photon fields}

The commutation relations with the photon field
yield
\be
[Q_{++},A_{+\mu}(x)]=2i\int\fr{d^3\bp}{(2\pi)^3}\fr{1}{N_{+}(\bp)}(b_{+}(\bp)e^{-ip_{+}.x}-ib^{\dag}_{+}(\bp)e^{ip_{+}.x})
                                  (e_{1\mu}-\fr{e_1.p_{+}}{l.p_{+}}l_{\mu})u_{+}(p_{+}),
\ee
which can be put in the form
\be
[Q_{++},A_{+\mu}(x)]=2i\psi_{+}(x)e_{1\mu}+2l_{\mu}(e_1.\d)\Om_{+}(x),
\ee
where the additional field $\Om_{+}(x)$ depends on the photon mode operators in the form
\be
\Om_{+}(x)=\int\fr{d^3\bp}{(2\pi)^3}\fr{1}{N_{+}(\bp)}\fr{1}{l.p_{+}}(b_{+}(\bp)e^{-ip_{+}.x}+ib^{\dag}_{+}(\bp)e^{ip_{+}.x})u_{+}(p_{+}),
\label{NLOC1}
\ee
and $\Om_{+}(x)$ satisfies
\be
il.\d\Om_{+}(x)=\psi_{+}(x).
\label{NLOC2}
\ee

We have also
\be
[Q_{--},A_{-\mu}(x)]=2\int\fr{d^3\bp}{(2\pi)^3}\fr{1}{N_{-}(\bp)}(b_{-}(\bp)e^{-ip_{-}.x}+ib^{\dag}_{-}(\bp)e^{ip_{-}.x})
				   (e_{2\mu}-\fr{e_2.p_{-}}{l.p_{-}}l_{\mu})u_{-}(p_{-}),
\ee
leading to
\be
[Q_{--},A_{-\mu}(x)]=2\psi_{-}(x)e_{2\mu}-2il_{\mu}(e_2.\d)\Om_{-}(x),
\ee
where
\be
\Om_{-}(x)=\int\fr{d^3\bp}{(2\pi)^3}\fr{1}{N_{-}(\bp)}\fr{1}{l.p_{-}}(b_{-}(\bp)e^{-ip_{-}.x}-ib^{\dag}_{-}(\bp)e^{ip_{-}.x})u_{-}(p_{-}),
\label{NLOC3}
\ee
and $\Om_{-}(x)$ satisfies
\be
il.\d\Om_{-}(x)=\psi_{-}(x).
\label{NLOC4}
\ee
It is interesting to note that the solutions of eq(\ref{NLOC2}) and eq(\ref{NLOC4}) can be expressed in the form
\be
\Om_{\pm}(x)=i\int_{0}^{\infty}ds e^{-\eps s}\psi_{\pm}(x+s l).
\label{NLOC5})
\ee
The limit $\eps\rightarrow 0+$ is assumed. The integration over $s$ creates the factor $1/(l.p_{\pm})$
in the integrands on the right sides of eq(\ref{NLOC1}) and eq(\ref{NLOC3}). The point here is that although, in this form, 
the right side of eq(\ref{NLOC5}) can be expressed directly in terms of the Majorana field, the result is not local
but requires an integration over a line of points emerging from $x$ in the light-like direction $l$. A similar factor
appeared in the integrand of the right side of eq(\ref{ANTICOM3}) for the anticommutators of $Q_{\pm}$.
This suggests that a nonlocal structure is intrinsic to the supersymmetry algebra that can be formulated 
in the presence of Lorentz symmetry breaking (of the type we have considered). This may help to explain why there 
seems to be no obvious Noether method for constructing the the supersymmetry charges in our model. 
In this context we should recall that while this is not true for an {\it extrinsic} Lorentz symmetry violation
our model has a violation that is {\it intrinsic}.

Finally we note that the fields $\phi_{\pm}(x)$ and $\Om_{\pm}(X)$ are related through the commutation relations
\be
[Q_{\pm\pm},\phi_{\pm}]=-\Om_{\pm}(x),
\ee
and
\be
	\{\Om_{+}(x),\Qb_{++}\}=i\int_{0}^{\infty} ds e^{-\eps s}\{\psi_{+}(x+s l),\Qb_{++}\},
\ee
and similarly
\be 
        \{\Om_{-}(x),\Qb_{--}\}=i\int_{0}^{\infty} ds e^{-\eps s} \{\psi_{-}(x+s l),\Qb_{--}\}.
\ee
Again we note the non-local character of these results. 

\subsection{\label{LILIM}Algebra of Supersymmetric Charges and Fields in the Lorentz Invariant Limit}

We again consider the limit of Lorentz symmetry. The four supersymmetry charges are given in eq(\ref{LISSQ}).
The non-vanishing anticommutators of the charges with the fields become
\begin{eqnarray}
	\{\psi_{+}(x),\Qb_{++}\}&=&\fr{1}{2}F_{+\ll\tau}\ss^{\ll\tau}-\gg.\d l.\d\phi_{+}(x)\nonumber\\
	\{\psi_{+}(x),\Qb_{-+}\}&=&\fr{1}{2}F_{-\ll\tau}\ss^{\ll\tau}+i\gg_{5}\gg.\d l.\d\phi_{-}(x)\nonumber\\
	\{\psi_{-}(x),\Qb_{--}\}&=&\fr{1}{2}F_{-\ll\tau}\ss^{\ll\tau}-i\gg_{5}\gg.\d l.\d\phi_{-}(x)\nonumber\\
	\{\psi_{-}(x),\Qb_{+-}\}&=&\fr{1}{2}F_{+\ll\tau}\ss^{\ll\tau}+\gg.\d l.\d\phi_{+}(x).
\end{eqnarray}
Using the definition in eq(\ref{TOTQ}) for the total supersymmetry charge $Q$ we can deduce that
\be
\{\psi(x),\Qb\}=F_{\ll\tau}\ss^{\ll\tau},
\ee
where, of course
\be
\psi(x)=\psi_{+}(x)+\psi_{-}(x),
\ee
and
\be
F_{\ll\tau}=F_{+\ll\tau}+F_{-\ll\tau}(x).
\ee
This is exactly what we expect for this Majorana field anticommutator in the Lorentz symmetric case.

The non-vanishing commutators for the photon fields become in the Lorentz invariant limit
is most conveniently considered in two stages. In the first stage we find directly
\begin{eqnarray}
	{[Q_{++},A_{+\mu}(x)]}&=&2i\psi_{+}(x)e_{1\mu}+2il_{\mu}e_{1}.\d\int_{0}^{\infty} ds e^{-\eps s}\psi_{+}(x+sl)\nonumber\\
	{[Q_{+-},A_{+\mu}(x)]}&=&-2i\psi_{-}(x)e_{1\mu}-2il_{\mu}e_{1}.\d\int_{0}^{\infty} ds e^{-\eps s}\psi_{-}(x+sl)\nonumber\\
	{[Q_{--},A_{-\mu}(x)]}&=&2\gg_5 \psi_{-}(x)e_{2\mu}+2l_{\mu}\gg_5e_{2}.\d\int_{0}^{\infty} ds e^{-\eps s}\psi_{-}(x+sl)\nonumber\\
	{[Q_{-+},A_{-\mu}(x)]}&=&-2\gg_5\psi_{+}(x)e_{2\mu}-2l_{\mu}\gg_5 e_{2}.\d\int_{0}^{\infty} ds e^{-\eps s}\psi_{+}(x+sl).
\end{eqnarray}
These equations, together with the related vanishing anticommutators, imply that

\begin{eqnarray}
	[Q,A_{\mu}(x)]&=&2(ie_{1\mu}-\gg_5 e_{2\mu})(\psi_{+}(x)-\psi_{-}(x))\nonumber\\
			       && +2l_{\mu}(ie_{1}-\gg_5e_{2}).\d\int_{0}^{\infty}dse^{-\eps s}(\psi_{+}(x+sl)-\psi_{-}(x+sl)),
\label{PHCOM}
\end{eqnarray}
where $Q$ is the total supersymmetry charge (see eq(\ref{TOTQ})) and
$A_{\mu}(x)=A_{+\mu}(x)+A_{-\mu}(x)$. We now make use of the identity
\be
\gg_{\mu}u_{+}(p)=2p_{\mu}u_{+}^{(0)}-l_{\mu}u_{n+}(p)+(ie_{1\mu}-\gg_5e_{2\mu})u_{+}(p),
\label{GGPID1}
\ee
where 
\be
u_{n+}(p)=\gg.p n.\gg u_{+}^{(0)}.
\ee
From eq(\ref{GGPID1}) we then obtain
\be
\gg_{\mu}\psi_{+}(x)=2\d_{\mu}\psi_{+}^{(0)}(x)-l_{\mu}\psi_{n+}(x)+(ie_{1\mu}-\gg_5 e_{2\mu})\psi_{+}(x).
\label{GGPID2}
\ee
We have for convenience, introduced
\be
\psi_{+}^{(0)}(x)=i\int\fr{d^3\bp}{(2\pi)^3}\fr{1}{N(\bp)}(b_{+}(\bp)e^{-ip.x}+ib_{+}^{\dag}(\bp)e^{ip.x})u_{+}^{(0)},
\label{GGPID3}
\ee
and
\be
\psi_{n+}(x)=\int\fr{d^3\bp}{(2\pi)^3}\fr{1}{N(\bp)}(b_{+}(\bp)e^{-ip.x}-ib_{+}^{\dag}(\bp)e^{ip.x})u_{n+}(p).
\label{GGPID4}
\ee
By multiplying the terms in eq(\ref{GGPID1}) by $\gg_5$ we obtain 
\be
-\gg_{\mu}u_{-}(p)=-2p_{\mu}u_{-}^{(0)}+l_{\mu}u_{n-}(p)+(e_{1\mu}-\gg_5e_{2\mu})u_{-}(p),
\label{GGPID5}
\ee
where $u_{-}^{(0)}=-\gg_5u_{+}^{(0)}$ and $u_{n-}(p)=-\gg_5u_{n+}(p)$.
This leads to
\be
-\gg_{\mu}\psi_{-}(x)=-2i\d_{\mu}\psi_{-}^{(0)}(x)+l_{\mu}\psi_{n-}(x)+(ie_{1\mu}-\gg_5e_{2\mu})\psi_{-}(x).
\label{GGPID6}
\ee
Here 
\be
\psi_{-}^{(0)}(x)=i\int\fr{d^3\bp}{(2\pi)^3}\fr{1}{N(\bp)}(b_{-}(\bp)e^{-ip.x}-ib^{\dag}_{-}(\bp)e^{ip.x})u_{-}^{(0)},
\ee
and 
\be
\psi_{n-}(x)=\int\fr{d^3\bp}{(2\pi)^3}\fr{1}{N(\bp)}(b_{-}(\bp)e^{-ip.x}+ib^{\dag}_{-}(\bp)e^{ip.x})u_{n-}(p).
\ee
Subtracting eq(\ref{GGPID6}) from eq(\ref{GGPID2}) we obtain
\be
\gg_{\mu}\psi(x)=2i\d_{\mu}\psi^{(0)}(x)-l_{\mu}\psi_{n}(x)+(ie_{1\mu}-\gg_5e_{2\mu})(\psi_{+}(x)-\psi_{-}(x)).
\label{GGPID7}
\ee
Here $\psi^{(0)}=\psi^{(0)}_{+}(x)+\psi^{(0)}_{-}(x)$ {\it etc}. We can use eq(\ref{GGPID7}) to eliminate 
$(\psi_{+}(x)-\psi_{-}(x))$ from eq(\ref{PHCOM}) with the result
\begin{eqnarray}
	[Q,A_{\mu}(x)]&=&2\gg_{\mu}\psi(x)-4\d_{\mu}\psi^{(0)}(x)+2l_{\mu}\psi_{n}(x)\nonumber\\
		    && +2l_{\mu}\d^{\aa}\int_{0}^{\infty} ds e^{-\eps s}(\gg_{\aa}\psi(x+sl)-2\d_{\aa}\psi^{(0)}(x)+l_{\aa}\psi_{n}(x+sl)).
\end{eqnarray}
We note that $\gg.\d\psi(x)=\d^2\psi^{(0)}(x)=0$ and 
\be
l.\d\int ds e^{-\eps s}\psi_{n}(x+sl)=-\psi_{n}(x),
\ee
we find
\be
[Q,A_{\mu}(x)]=2\gg_{\mu}\psi(x)-4\d_{\mu}\psi^{(0)}(x).
\label{MACOMM}
\ee
The  outcome is exactly what we expect for the Lorentz symmetric model apart from
the derivative term on the right side of eq(\ref{MACOMM}). This can be accommodated 
by incorporating an appropriate gauge transformation of $A_{\mu}(x)$ and in any case
does not affect the result for $[Q,F_{\mu\nu}(x)]$. In fact we could have approached the
whole analysis in the Lorentz symmetric case along the lines we have set out above without
reference to Lorentz symmetry breaking. It would of course seem a rather roundabout approach.
The key point is that when we have Lorentz symmetry there are {\it four} conserved
supersymmetry charges that may be combined to make up the total supersymmetry charge.
Whereas there are only {\it two} when we violate Lorentz symmetry in the manner
of our model. 

\section{Wess-Zumino Model}

It is interesting to compare the results of the photon/photino model with similar birefringent phenomena
that can be obtained in a simple non-interecting Wess-Zumino model \cite{WESSZUM} with the Majorana photino field and two scalar fields.
We retain the Majorana Lagrangian of eq(\ref{MAJMAJL}). We introduce a scalar field $\Phi_{+}(x)$ and a
pseudo scalar field $\Phi_{-}(x)$ with a Lagrangian $\Lc_S(x)$ where
\be
\Lc_S(x)=\fr{1}{2}((\d+(\kk/2)l~l.\d)\Phi_{+}(x))^2+\fr{1}{2}((\d-(\kk/2)l~l.\d)\Phi_{-}(x))^2.
\label{SCASCAL}
\ee
The fields $\Phi_{\pm}(x)$ satisfy the equations of motion
\be
(\d^2\pm\kk(l.\d)^2)\Phi_{\pm}(x)=0.
\ee
The plane waves satisfying these wave equations are
\be
\Phi_{\pm}(x)=e^{-ip_{\pm}.x},
\ee
where
\be
p_{\pm}^2\pm\kk(l.p_{\pm})^2=0.
\ee
It is obvious that these dispersion relations for the scalar particles associated with
these plane waves yield the same "$\pm$"lightcones as for the photons and the photinos.

The quantum fields can be expanded in terms of these wave functions in the form
\be
\Phi_{\pm}(x)=\int\fr{d^3\bp}{(2\pi)^2}\fr{1}{\Nc_{S\pm}(\bp)}(c_{\pm}(\bp)e^{-ip_{\pm}.x}+c^{\dag}_{\pm}(\bp)e^{ip_{\pm}.x}),
\ee
where
\be
[c_{\pm}(\bp),c^{\dag}_{\pm}(\bp')]=(2\pi)^3\dd(\bp-\bp'),
\ee
and $\Nc_{S\pm}(\bp)$ is chosen to guarantee the canonical commutation relations, that is 
\be
\Nc_{S\pm}=\sqrt{2(E_{\pm}\pm(\kk/2)(E_{\pm}-p_3))}
\ee

\subsection{\label{WZSSCS}Supersymmetry Charges with Scalar Fields}

Because of the conformity of the scalar and photino lightcones we
can realise the associated supersymmetry by means of conserved currents
$J^{\mu}_{S\pm\pm}(x)$ that give rise to conserved supersymmetry
charges $Q_{S\pm\pm}$, where
\be
Q_{S\pm\pm}=\int d^3\bx J^0_{S\pm\pm}(x).
\ee
First we note the modified currents 
\be
	\Jht^{\mu}_{S++}(x)=i\dht_{+\ll}\Phi_{+}(x)\gg^{\ll}\gg^{\mu}\psi_{+}(x),
\label{SCSSC1}
\ee
and
\be
	\Jht^{\mu}_{S--}(x)=\gg_5\dht_{-\ll}\Phi_{-}(x)\gg^{\ll}\gg^{\mu}\psi_{-}(x),
\label{SCSSC2}
\ee
satisfy
\be
\dht_{\pm\mu}\Jht^{\mu}_{S\pm\pm}(x)=0.
\ee
The factor $\gg_5$ in eq(\ref{SCSSC2}) is for future convenience. It relates to the fact that
$\Phi_{+}(x)$ and $\Phi_{-}(x)$ have opposite parities.
We then set
\be
J^{\mu}_{S\pm\pm}(x)=\Jht^{\mu}_{S\pm\pm}(x)\pm(\kk/2)l^{\mu}l.\Jht_{S\pm\pm}(x).
\ee
It follow from the equations of motion, that 
\be
\d_{\mu}J^{\mu}_{S\pm\pm}(x)=0.
\ee
The associated supersymmetry charges are then also conserved.
We find
\be
Q_{S++}=\int\fr{d^3\bp}{(2\pi)^3}\fr{1}{\sqrt{(E_{+}-p_3)}}(-ic_{+}(\bp)b^{\dag}_{+}(\bp)-c^{\dag}_{+}(\bp)b_{+}(\bp))u_{+}(p_{+}).
\ee
\be
Q_{S--}=\int\fr{d^3\bp}{(2\pi)^3}\fr{1}{\sqrt{(E_{-}-p_3)}}(c_{-}(\bp)b^{\dag}_{-}(\bp)+ic^{\dag}_{-}(\bp)b_{-}(\bp))u_{+}(p_{-}).
\ee

These have the same form (up to a normalisation) as the corresponding charges linking photons and photinos. The nonvanishing anticommutation 
relations are
\be
\{Q_{S\pm\pm},\Qb_{S\pm\pm}\}=2\int\fr{d^3\bp}{(2\pi)^3}\fr{1}{(E_{\pm}-p_3)}(c^{\dag}_{\pm}(\bp)c_{\pm}(\bp)+b^{\dag}_{\pm}(\bp)b_{\pm}(\bp))u_{+}(p_{\pm})\ub_{+}(p_{\pm}).
\ee
Making use of the identity in eq(\ref{WVFNID}) we have
\be
\{Q_{S++},\Qb_{S++}\}=\int\fr{d^3\bp}{(2\pi)^3}(c^{\dag}_{+}(\bp)c_{+}(\bp)+b^{\dag}_{+}(\bp)b_{+}(\bp))
				   \left[\gg.\pht_{+}+\rt2\fr{\veps^{\ll}_{+}(p_{+})\pht_{+}^{\tau}\ss_{\ll\tau}}{(E_{+}-p_3)}\right],
\ee
and
\be
\{Q_{S--},\Qb_{S--}\}=\int\fr{d^3\bp}{(2\pi)^3}(c^{\dag}_{-}(\bp)c_{-}(\bp)+b^{\dag}_{-}(\bp)b_{-}(\bp))
				   \left[\gg.\pht_{-}+\rt2\fr{\veps^{\ll}_{+}(p_{-})\pht_{-}^{\tau}\ss_{\ll\tau}}{(E_{-}-p_3)}\right].
\ee

\subsection{\label{WZCOMMS}Commutation Relations of Supersymmetry Charges with Fields}

We have for the mode operators
\begin{eqnarray}
	[Q_{S++},c_{+}(\bp)]&=&\fr{1}{\sqrt{(E_{+}-p_3)}}b_{+}(\bp)u_{+}(p_{+})\nonumber\\
	{[Q_{S++},c^{\dag}_{+}(\bp)]}&=&\fr{-i}{\sqrt{(E_{+}-p_3)}}b^{\dag}_{+}(\bp)u_{+}(p_{+})\nonumber\\
	\{b_{+}(\bp),\Qb_{S++}\}&=&\fr{-1}{\sqrt{(E_{+}-p_3)}}c_{+}(\bp)\ub_{+}(p_{+})\nonumber\\
	\{b^{\dag}_{+}(\bp),\Qb_{S++}\}&=&\fr{i}{\sqrt{(E_{+}-p_3)}}c^{\dag}_{+}(\bp)\ub_{+}(p_{+}),
\end{eqnarray}
and
\begin{eqnarray}
        [Q_{S--},c_{-}(\bp)]&=&\fr{-i}{\sqrt{(E_{-}-p_3)}}b_{-}(\bp)u_{+}(p_{-})\nonumber\\
	{[Q_{S--},c^{\dag}_{-}(\bp)]}&=&\fr{1}{\sqrt{(E_{-}-p_3)}}b^{\dag}_{-}(\bp)u_{+}(p_{-})\nonumber\\
	\{b_{-}(\bp),\Qb_{S--}\}&=&\fr{-i}{\sqrt{(E_{-}-p_3)}}c_{-}(\bp)\ub_{+}(p_{-})\nonumber\\
	\{b^{\dag}_{-}(\bp),\Qb_{S--}\}&=&\fr{1}{\sqrt{(E_{-}-p_3)}}c^{\dag}_{-}(\bp)\ub_{+}(p_{-}),
\end{eqnarray}
giving rise to the results
\begin{eqnarray}
	[Q_{S++},\Phi_{+}(x)]&=&\psi_{+}(x)\nonumber\\
	\{\psi_{+}(x),\Qb_{S++}\}&=&\fr{-i}{2}[\gg.\dht_{+}\Phi_{+}(x)+e_{1\ll}\dht_{\tau}\ss^{\ll\tau}\Phi_{+}(x)
			       +e_{1}.\d l_{\ll}\dht_{+\tau}\ss^{\ll\tau} \int_{0}^{\infty}dse^{-\eps s}\Phi_{+}(x+sl)]\nonumber\\
			       {}&&{}
\label{SFCOMM1}			       
\end{eqnarray}
and
\begin{eqnarray}
	[Q_{S--},\Phi_{-}(x)]&=&-i\gg_5\psi_{-}(x)\nonumber\\
	\{\psi_{-}(x),\Qb_{S--}\}&=&\fr{1}{2}\gg_5[\gg.\dht_{-}\Phi_{-}(x)+e_{1\ll}\dht_{-\tau}\ss^{\ll\tau}\Phi_{-}(x)
				+e_{1}.\d l_{\ll}\dht_{-\tau}\ss^{\ll\tau}\int_{0}^{\infty}ds e^{-\eps s}\Phi_{-}(x+sl)]\nonumber\\
				{}&&{}
\label{SFCOMM2}		
\end{eqnarray}
These results are similar in character to those for the photon/photino model and reveal the presence of
nonlocal terms in the anticommutators of the conserved supersymmetry charges with the photino fields.
The commutators with the scalar fields are in this case purely local and yield the photino fields without
any further contributions.

\subsection{\label{INTCCWZ} Interchange Mass-Shell Cones in Wess-Zumino Model}

Just as for the photon/photino model we can interchange the mass-shell cones of the Majorana 
particles by setting $\xi=-\kk/(2\rt2)$. This leads in a straightforward way to conserved
supercurrents $J_{S\pm\mp}^{\mu}(x)=\Jht_{S\pm\mp}(x)\pm(\kk/2)l^{\mu}l.\Jht^{\mu}_{S\pm\mp}(x)$ where
\be
\Jht^{\mu}_{S+-}(x)=i\dht_{+\ll}\Phi_{+}(x)\gg^{\ll}\gg^{\mu}\psi_{-}(x),
\label{SCSSC3}
\ee
and
\be
\Jht^{\mu}_{S-+}(x)=\gg_5\dht_{-\ll}\Phi_{-}(x)\gg^{\ll}\gg^{\mu}\psi_{+}(x).
\label{SCSSC4}
\ee
Here $\psi_{\mp}(x)$ has the form given by eq(\ref{INTRCHa}). Computing the now conserved supercharges
in the usual way we find
\be
Q_{S+-}=\int\fr{d^3\bp}{(2\pi)^3}\fr{1}{\sqrt{(E_{+}-p_3)}}( ic_{+}(\bp)b^{\dag}_{-}(\bp)
		    -c^{\dag}_{+}(\bp)b_{-}(\bp))u_{-}(p_{+}).
\ee
\be
Q_{S-+}=\int\fr{d^3\bp}{(2\pi)^3}\fr{1}{\sqrt{(E_{-}-p_3)}}(-c_{-}(\bp)b^{\dag}_{+}(\bp)
                    +ic^{\dag}_{-}(\bp)b_{+}(\bp))u_{-}(p_{-}).
\ee

We have, following the same pattern of argument as subsection(\ref{INTCGCNS})
\be
\{Q_{S\pm\mp},\Qb_{S\pm\mp}\}=\int\fr{d^3\bp}{(2\pi)^3}
		 \left[\gg.\pht_{\pm}-\fr{\rt2\veps_{\pm\ll}(p_{\pm})\pht_{\pm\tau}\ss^{\ll\tau}}{E_{\pm}-p_3}\right]
		 (c^{\dag}_{\pm}(\bp)c_{\pm}(\bp)+b^{\dag}_{\mp}(\bp)b_{\mp}(\bp)).
\ee
The non-vanishing (anti)commutation relations of these charges are
\begin{eqnarray}
[Q_{S+-},\Phi_{+}(x)]&=&\psi_{-}(x)\nonumber\\
	\{\psi_{-}(x),\Qb_{S+-}\}&=&\fr{-i}{2}[\gg.\dht_{+}\Phi_{+}(x)-e_{1\ll}\dht_{+\tau}\ss^{\ll\tau}\Phi_{+}(x)
		-e_1.\d l_{\ll}\dht_{+\tau}\ss^{\ll\tau}\int_0^{\infty} dse^{-\eps  s}\Phi_{+}(x+sl)]\nonumber\\
                               {}&&{}
\label{SFCOMM3}
\end{eqnarray}
and
\begin{eqnarray}
	[Q_{-+},\Phi_{-}(x)]&=&-i\gg_5\psi_{+}(x)\nonumber\\
	\{\psi_{+}(x),\Qb_{S-+}\}&=&\fr{1}{2}\gg_5[\gg.\dht_{-}\Phi_{-}(x)-e_{1\ll}\dht_{-\tau}\ss^{\ll\tau}\Phi_{-}(x)
		  -e_{1}.\d l_{\ll}\dht_{-\tau}\ss^{\ll\tau}\int_0^{\infty} ds e^{-\eps s}\Phi_{-}(x+sl) ]\nonumber\\
			      {}&&{}
\label{SFCOMM4}
\end{eqnarray}
The results in eq(\ref{SFCOMM3}) and eq(\ref{SFCOMM4}), although a little simpler than the corresponding
results for the photon/photino model, have in common the feature of involving nonlocal terms with integrals over a line
starting at $x$ and parallel to the null 4-vector $l$.

\subsection{\label{LILWZ} Lorentz Invariant Limit for the Wess-Zumino Model}

In the same way as for the photon/photino model we can examine the limit of Lorentz invariance for the 
Wess-Zumino model. Recall that when $\kk\rightarrow 0\pm$ the 4-momenta satisfy $p_{\pm}=\pht_{\pm}=p$ where
$p^2=0$ and all of the supercharges are conserved. The form of the of the supersymmetry charges is 
listed in appendix(\ref{WZLSL}) together with the complete set of anticommutators. We again find that the
off-diagonal anticommutators cancel in pairs. We may therefore define an overall supersymmetry charge 
$Q_{S}=Q_{S++}+Q_{S--}+Q_{S+-}+Q_{S-+}$. It is easy to establish that
\be
\{Q_{S},\Qb_{S}\}=2\gg.P,
\ee
where $P^{\mu}$ is the 4-momentum operator for the fields in the model. 

From eqs(\ref{SFCOMM1},\ref{SFCOMM2},\ref{SFCOMM3},\ref{SFCOMM4}) we can, in the Lorentz invariant limit,
obtain the results
\begin{eqnarray}
	{[Q_{S++},\Phi_{+}(x)]}&=&\psi_{+}(x)\nonumber\\
	{[Q_{S--},\Phi_{-}(x)]}&=&-i\gg_5\psi_{-}(x)\nonumber\\
	{[Q_{S+-},\Phi_{+}(x)]}&=&\psi_{-}(x)\nonumber\\
	{[Q_{S-+},\Phi_{-}(x)]}&=&-i\gg_5\psi_{+}(x)
	\label{SFCOMM5}
\end{eqnarray}
These may be combined appropriately to yield
\be
[Q_{S},\Phi_{+}(x)\pm i\Phi_{-}(x)]=(1\pm\gg_5)\psi(x).
\ee
We can also extract the results
\begin{eqnarray}
	\{\psi_{+}(x),\Qb_{S++}\}&=&-\fr{i}{2}\gg.\d\Phi_{+}(x)+...\nonumber\\
	\{\psi_{-}(x),\Qb_{S--}\}&=&\fr{1}{2}\gg_5\gg.\d\Phi_{-}(x)+...\nonumber\\
	\{\psi_{-}(x),\Qb_{S+-}\}&=&-\fr{i}{2}\gg.\d\Phi_{+}(x)+...\nonumber\\
	\{\psi_{+}(x),\Qb_{S-+}\}&=&\fr{1}{2}\gg_5\gg.\d\Phi_{-}(x)+...
	\label{SFCOMM6}
\end{eqnarray}
The ellipses in eqs(\ref{SFCOMM6}) indicate terms that cancel from the final result. In particular these 
cancelling terms include the nonlocal contributions.
Finally we obtain the purely local result
\be
\{\psi(x),\Qb_{S}\}=-i\gg.\d\Phi_{+}(x)+\gg_5\gg.\d\Phi_{-}(x).
\ee
This result may be re-expressed as
\be
\{(1\pm\gg_5)\psi(x),\Qb_{S}\}=\pm i(1\pm\gg_5)\gg.\d(\Phi_{+}(x)\pm i\Phi_{-}(x)),
\ee
which is the standard result for the Lorentz invariant Wess-Zumino model.

\section{\label{CONC}Conclusions}

We have argued for a distinction between an {\it extrinsic} breaking of Lorentz 
symmetry that can be "removed" by an appropriate coordinate transformation
and an {\it intrinsic} breaking that cannot be so removed. In the extrinsic case 
such a model can retain in full the original supersymmetry in a "disguised" form
but with essentially the same formal algebraic structure. This does not mean that the violation 
of Lorentz invariance is illusory but merely that the allowed obsever frames
may differ from the coordinate frames natural to the model under investigation. 
In the intrinsic case the origin of Lorentz symmetry violation is irremovable
by a change of coordinates. A clear case is the presence of birefringence. 
It is then less obvious {\it a priori} that any supersymmetry
can be retained. However we have shown from an examination of two simple models,
a photon/photino model and a Wess-Zumino model, both exhibiting birefringence,
that some supersymmetry can indeed be retained. 

In the photon/photino model the Lorentz symmetry violation for the photons is provided by coupling the vector field
to a Weyl tensor background field. The Weyl tensor is of class N in the Petrov classification
scheme. This is the the simplest case. The result is a birefringent form of photon propagation.
The photino field is a (massless) Majorana field coupled to a Lorentz symmetry violating background
bi-vector field related to the (Petrov class N) Weyl field controlling photon propagation. By adjusting the 
strength of the coupling it is possible to align the birefringence of the photinos with that
of the photons. This means that for every photon of spatial momentum $\bp$ there is an
appropriate companion photino with the same momentum and energy. This is the basis of the supersymmetry
in the model. In the Wess-Zumino model we retain the Majorana field and adjust the scalar field 
propagation so that it exhibits birefringence that matches appropriately the propagation of the Majorana field.

Consideration of the double lightcone structure that governs the propagation of both photons 
and photinos or scalar particles and Majorana particles, 
shows that there are limitations on the ensemble of allowed observational reference
frames. No boost to a reference frame is acceptable if it requires the observer to travel faster
than particles on the slower lightcone. This constraint can be re-expressed as a bound on the 
magnitude of the coupling to the Lorentz violating background fields. When this bound is broken the 
(positive time) slow cone tilts over so far that the time axis of the observer's coordinate system
no longer lies within it thus rendering the coordinates inappropriate for describing the unitary 
evolution of the model. Of course in order to genuinely observe the system there has to be an 
interrogating interaction between model and observer. In that case an
observer breaking the slow cone speed barrier will generate an associated shower of 
\v{C}erenkov radiation (see refs \cite{LEPOT,SCHR3}) thus paying a speeding penalty by ceasing to be an observer and
effectively becoming part of the model. This \v{C}erenkov phenomenon can of course be 
detected by observers that are obeying the speed limit. These considerations apply more generally 
to any model with multiple lightcones.

If we accept the speeding restriction on our observation frame we can proceed to study the 
equations of motion of the model and quantise it in a conventional way. We use a slightly
modified version of the Gupta-Bleuler method to quantise the the vector field $A_{\mu}(x)$ 
of the photon resulting in a breakup into three parts, $A_{+\mu}(x)$, associated with
the slow cone, $A_{-\mu}(x)$ associated with the fast cone and $A_{U\mu}(x)$
which comprises a pure derivative guage field and a conjugate zero norm term. They do not
contribute to matrix elements of physical observables. Interestingly the lightcone appropriate 
for the propagation of $A_{U\mu}(x)$, is the standard cone that is invariant under the
Lorentz transformations that connect the coordinate systems of the set of observers.
For this additional reason it is essential that $A_{U\mu}(x)$ is not involved in the
construction of observables which relate to the energies and momenta of photons and photinos
travelling on the fast and slow lightcones. 

The Majorana equation appropriate to the photino field $\psi(x)$
is modified in a way that breaks Lorentz invariance. The resulting birefringence, 
makes it possible to view it as the sum of two parts, $\psi_{+}(x)$ associated with
the slow cone and $\psi_{-}(x)$ associated with the fast cone. 
From $A_{+\mu}(x)$ and $\psi_{+}(x)$ we can construct a (spinor-valued) current $J_{++\mu}(x)$
that is conserved and which gives rise to a constant supersymmetry charge $Q_{++}$.
Similarly a supersymmetry charge $Q_{--}$ can be obtained from a conserved current
$J_{--\mu}(x)$ constructed from $A_{-\mu}(x)$ and $\psi_{-}(x)$. The non-zero anticommutation
relations satisfied by the two charges each have two contributions (see eq(\ref{ANTICOM3})).
One part has the form $\gg.\Pht_{\pm}$ and is analogous to the standard result
but with the the momentum operator $P_{\mu}$ replaced by $\Pht_{\pm\mu}$. The second part
has a more elaborate form that we argue later has indications of non-locality.
In the Lorentz invariant limit ($\kk=0$) the first part reduces to the standard form
but the second part remains. The full understanding of this limit requires an examination 
of the regime in which $\xi=-\kk/2\rt2$. In this case the the fast and slow lightcones
for photinos interchange and the conserved supersymmetry currents are 
$J_{+-\mu}(x)$ and $J_{-+\mu}(x)$ (see eq(\ref{ANTICOM4})). When $\kk=0$ all four currents are conserved
and it is then possible to build a total supersymmetry charge that has the standard
anticommutation relations (see section \ref{LSYMLIM}). It is evident then that the limit back 
to Lorentz invariance is not straightforward and issues related to the complexity of the algebra of 
conserved charges reappear when their effect on the photon and photino fields are considered.

The various (anti)commutation relations with the mode operators is straightforward and the 
conversion $a_{\pm}(\bp)\leftrightarrow b_{\pm}(\bp)$ is as expected (see eq(\ref{MODCOMS1}) and eq(\ref{MODCOMS2})).
The (anti)commutation relations of $Q_{\pm\pm}$ with the dynamical fields is rather less conventional
and involves the introduction of the modified fields $\phi_{\pm}(x)$ and $\Om_{\pm}(x)$ that have a non-local
relationship with $A_{\pm\mu}(x)$ and $\psi_{\pm}(x)$ even though they are constructed from the same sets of 
mode operators. We speculate that this non-local property is related to the apparent impossibility of constructing
in our photon/photino model, conserved supersymmetry currents by means of the Noether method. Nevertheless
if we accept these complexities and limitations of the model we can argue that even in this model with intrinsic 
Lorentz symmetry violation it is possible to identify a remaining supersymmetry structure.

The Wess-Zumino model with Lorentz symmetry breaking exhibits the same features as the
photon/photino model. There is the same reduced number of conserved supersymmetry 
currents and charges. The (anti)commutation relations with the fields again show non-local
outcomes. The limit back to Lorentz symmetry yields a second set of of conserved charges that
permit the construction of a complete conserved charge $Q_{S}$ with the appropriate algebraic
properties for the Lorentz invariant Wess-Zumino model. The non-local terms again cancel
in a satifactory manner.

Of course a major limitation of our models is that the fields and their associated particles
are non-interacting. It would therefore be of interest to generalise the models so that the fields 
do have interactions such as a non-abelian gauge invariance. The task would, in particular, be to find a method
of identifying $A_{\pm a\mu}(x)$, $\psi_{\pm a}(x)$ and $\Phi_{\pm a}(x)$ ($a$ being the group multiplet label). 
This would presumably be achieved perturbatively and would intersect with the related task of  
confirming the renormalisation properties of the interacting model. It would also be interesting to 
generalise the nature of the Lorentz symmetry violation for the photon/photino model to other Petrov classes. In this context it
should be noted that Petrov class D is the only other example of the Weyl tensor giving rise to a dispersion
relation that factorises into two lightcones. The other classes, I, II and III of
Weyl tensor yield dispersion relations that are intrinsically quartic in the photon momentum \cite{ITD3}.
Since a product structure for the dispersion relation holds generally for modified Dirac equations \cite{COLL3}
this may prevent the aligning of the photon and photino dispersion relations and
the achieving of supersymmetry in the photon/photino model in those cases. 

Finally we remark that there is an intriguing parallelism of supersymmetry structure
between our models and the $N=1/2$ supersymmetry proposed by
Seiberg \cite{SEIB} in the context on noncommutative spacetime geometry. A major
difference between the models is that the Seiberg model is based on the standard chiral
structure of supersymmetry and retains Lorentz invariance whereas in the models analysed here the 
the violation of Lorentz invariance replaces chiral structure with charge conjugation symmetry.

\appendix

\section{\label{ACOMS}Anticommutators for Supersymmetry Charges in Lorentz Symmetry Limit}

We list here the anticommutators, in the Lorentz symmetry limit, for the supersymmetry charges of the photon/photino model.
The 4-momentum $p$ satisfies $p^2=0$. We have made the contractions $a_{\pm}(\bp)\rightarrow a_{\pm}$ {\it etc} for compactness.
\begin{eqnarray}
	\{Q_{++},\Qb_{++}\}&=&8\int\fr{d^3\bp}{(2\pi)^3}\fr{1}{E-p_3}(a^{\dag}_{+}a_{+}+b^{\dag}_{+}b_{+})u_{+}(p)\ub_{+}(p)\nonumber\\
	\{Q_{++},\Qb_{--}\}&=&0\nonumber\\
	\{Q_{++},\Qb_{+-}\}&=&4\int\fr{d^3\bp}{(2\pi)^3}\fr{1}{E-p_3}(b^{\dag}_{+}b_{-}+b_{+}b^{\dag}_{-})u_{+}(p)\ub_{-}(p) \nonumber\\
	\{Q_{++},\Qb_{-+}\}&=&-4i\int\fr{d^3\bp}{(2\pi)^3}\fr{1}{E-p_3}(a^{\dag}_{+}a_{-}+a_{+}a^{\dag}_{-})u_{+}(p)\ub_{-}(p)\nonumber\\
	\{Q_{--},\Qb_{--}\}&=&8\int\fr{d^3\bp}{(2\pi)^3}\fr{1}{E-p_3}(a^{\dag}_{-}a_{-}+b^{\dag}_{-}b_{-})u_{+}(p)\ub_{+}(p)\nonumber\\
	\{Q_{--},\Qb_{++}\}&=&0\nonumber\\
	\{Q_{--},\Qb_{+-}\}&=&4i\int\fr{d^3\bp}{(2\pi)^3}\fr{1}{E-p_3}(a^{\dag}_{-}a_{+}+a_{-}a^{\dag}_{+})u_{+}(p)\ub_{-}(p)\nonumber\\
	\{Q_{--},\Qb_{-+}\}&=&4\int\fr{d^3\bp}{(2\pi)^3}\fr{1}{E-p_3}(b^{\dag}_{-}b_{+}+b_{-}b^{\dag}_{+})u_{+}(p)\ub_{-}(p)\nonumber\\
	\{Q_{+-},\Qb_{+-}\}&=&8\int\fr{d^3\bp}{(2\pi)^3}\fr{1}{E-p_3}(a^{\dag}_{+}a_{+}+b^{\dag}_{-}b_{-})u_{-}(p)\ub_{+}(p)\nonumber\\
	\{Q_{+-},\Qb_{-+}\}&=&0\nonumber\\
	\{Q_{+-},\Qb_{++}\}&=&4\int\fr{d^3\bp}{(2\pi)^3}\fr{1}{E-p_3}(b^{\dag}_{-}b_{+}+b_{-}b^{\dag}_{+})u_{-}(p)\ub_{+}(p)\nonumber\\
	\{Q_{+-},\Qb_{--}\}&=&-4i\int\fr{d^3\bp}{(2\pi)^3}\fr{1}{E-p_3}(a^{\dag}_{+}a_{-}+a_{+}a^{\dag}_{-})u_{-}(p)\ub_{+}(p)\nonumber\\
	\{Q_{-+},\Qb_{-+}\}&=&8\int\fr{d^3\bp}{(2\pi)^3}\fr{1}{E-p_3}(a^{\dag}_{-}a_{-}+b^{\dag}_{+}b_{+})u_{-}(p)\ub_{-}(p)\nonumber\\
	\{Q_{-+},\Qb_{+-}\}&=&0\nonumber\\
	\{Q_{-+},\Qb_{++}\}&=&4i\int\fr{d^3\bp}{(2\pi)^3}\fr{1}{E-P_3}(a_{-}a^{\dag}_{+}+a^{\dag}_{-}a_{+})u_{-}(p)\ub_{+}(p)\nonumber\\
	\{Q_{-+},\Qb_{--}\}&=4&\int\fr{d^3\bp}{(2\pi)^3}\fr{1}{E-p_3}(b^{\dag}_{+}b_{-}+b_{+}b^{\dag}_{-})u_{-}(p)\ub_{+}(p)\nonumber
\end{eqnarray}
The cancelling pairs of off-diagonal contributions are obvious from the above list.

\section{\label{WZLSL} Lorentz Symmetric Limit for the Wess-Zumino Model}

Using the same notational conventions as in appendix(\ref{ACOMS}) we have, in the limit of Lorentz invariance,
\begin{eqnarray}
	Q_{S++}&=&\int\fr{d^3\bp}{(2\pi)^3}\fr{1}{\sqrt{E-p_3}}(-ic_{+}b^{\dag}_{+}-c^{\dag}_{+}b_{+})u_{+}(p)\nonumber\\
	Q_{S--}&=&\int\fr{d^3\bp}{(2\pi)^3}\fr{1}{\sqrt{E-p_3}}(c_{-}b^{\dag}_{-}+ic^{\dag}_{-}b_{-})u_{+}(p)\nonumber\\
	Q_{S+-}&=&\int\fr{d^3\bp}{(2\pi)^3}\fr{1}{\sqrt{E-p_3}}(ic_{+}b^{\dag}_{-}-c^{\dag}_{+}b_{-})u_{-}(p)\nonumber\\
	Q_{S-+}&=&\int\fr{d^3\bp}{(2\pi)^3}\fr{1}{\sqrt{E-p_3}}(-c_{-}b^{\dag}_{+}+ic^{\dag}_{-}b_{+})u_{-}(p)
\end{eqnarray}
The anticommutation relations for these charges are
\begin{eqnarray}
	\{Q_{S++},\Qb_{S++}\}&=&\int\fr{d^3\bp}{(2\pi)^3}\fr{2}{E-p_3}(c^{\dag}_{+}c_{+}+b^{\dag}_{+}b_{+})u_{+}(p)\ub_{+}(p)\nonumber\\
	\{Q_{S++},\Qb_{S--}\}&=&0\nonumber\\
	\{Q_{S++},\Qb_{S+-}\}&=&\int\fr{d^3\bp}{(2\pi)^3}\fr{1}{E-p_3}(-b^{\dag}_{+}b_{-}+b^{\dag}_{-}b_{+})u_{+}(p)\ub_{-}(p)\nonumber\\
	\{Q_{S++},\Qb_{S-+}\}&=&\int\fr{d^3\bp}{(2\pi)^3}\fr{i}{E-p_3}(c^{\dag}_{+}c_{-}+c^{\dag}_{+}c_{-})u_{+}(p)\ub_{-}(p)\nonumber\\
	\{Q_{S--},\Qb_{S++}\}&=&0\nonumber\\
	\{Q_{S--},\Qb_{S--}\}&=&\int\fr{d^3\bp}{(2\pi)^3}\fr{2}{E-p_3}(c^{\dag}_{-}c_{-}+b^{\dag}_{-}b_{-})u_+{}(p)\ub_{+}(p)\nonumber\\
	\{Q_{S--},\Qb_{S+-}\}&=&\int\fr{d^3\bp}{(2\pi)^3}\fr{-i}{E-p_3}(c^{\dag}_{+}c_{-}+c^{\dag}_{-}c_{+})u_{+}(p)\ub_{-}(p)\nonumber\\
	\{Q_{S--},\Qb_{S-+}\}&=&\int\fr{d^3\bp}{(2\pi)^3}\fr{1}{E-p_3}(-b^{\dag}_{-}b_{+}+b^{\dag}_{+}b_{-})u_{+}(p)\ub_{-}(p)\nonumber\\
	\{Q_{S+-},\Qb_{S++}\}&=&\int\fr{d^3\bp}{(2\pi)^3}\fr{1}{E-p_3}(-b^{\dag}_{-}b_{+}+b^{\dag}_{+}b_{-})u_{-}(p)\ub_{+}(p)\nonumber\\
	\{Q_{S+-},\Qb_{S--}\}&=&\int\fr{d^3\bp}{(2\pi)^3}\fr{i}{E-p_3}(c^{\dag}_{-}c_{+}+b^{\dag}_{+}b_{-})u_{-}(p)\ub_{+}(p)\nonumber\\
	\{Q_{S+-},\Qb_{S+-}\}&=&\int\fr{d^3\bp}{(2\pi)^3}\fr{2}{E-p_3}(c^{\dag}_{+}c_{+}+b^{\dag}_{-}b_{-})u_{-}(p)\ub_{-}(p)\nonumber\\
        \{Q_{S+-},\Qb_{S-+}\}&=&0\nonumber\\
	\{Q_{S-+},\Qb_{S++}\}&=&\int\fr{d^3\bp}{(2\pi)^3}\fr{-i}{E-p_3}(c^{\dag}_{+}c_{-}+c^{\dag}_{-}c_{+})u_{-}(p)\ub_{+}(p)\nonumber\\
	\{Q_{S-+},\Qb_{S--}\}&=&\int\fr{d^3\bp}{(2\pi)^3}\fr{1}{E-p_3}(-b^{\dag}_{+}b_{-}+b^{\dag}_{-}b_{+})u_{-}(p)\ub_{+}(p)\nonumber\\
	\{Q_{S-+},\Qb_{S+-}\}&=&0\nonumber\\
	\{Q_{S-+},\Qb_{S-+}\}&=&\int\fr{d^3\bp}{(2\pi)^3}\fr{2}{E-p_3}(c^{\dag}_{-}c_{-}+b^{\dag}_{+}b_{+})u_{-}(p)\ub_{-}(p)
\end{eqnarray}
By inspection it is clear that the off-diagonal anticommutators cancel in the sum.

\section*{Acknowledgements}

We acknowledge many helpful comments and suggestions by G. M. Shore and J. M. Drummond.
This work has been partially supported by STFC consolidated grant ST/T000694/1.

\bibliography{mm2n}
\bibliographystyle{unsrt}

\end{document}